\documentclass[showpacs,prl,onecolumn,aps,superscriptaddress,preprintnumbers,nofootinbib]{revtex4}
\usepackage[T1]{fontenc}
\usepackage[latin9]{inputenc}
\usepackage{graphicx,hyperref}

\usepackage{amsmath,amssymb}
\usepackage{epsfig}
\usepackage{graphicx}
\usepackage{amsmath}
\usepackage{amsfonts}
\usepackage{mathrsfs}
\usepackage{pifont}
\usepackage{relsize}
\usepackage{mathtools}
\def\slashchar#1{\setbox0=\hbox{$#1$}     		
   \dimen0=\wd0                                 	
   \setbox1=\hbox{/} \dimen1=\wd1               	
   \ifdim\dimen0>\dimen1                        	
      \rlap{\hbox to \dimen0{\hfil/\hfil}}      	
      #1                                        	
   \else                                        	
      \rlap{\hbox to \dimen1{\hfil$#1$\hfil}}   	
      /                                         	
   \fi}

\renewcommand{\vec}{\boldsymbol}
\newcommand{\beq}{\begin{equation}}
\newcommand{\eeq}{\end{equation}}
\newcommand{\bea}{\begin{eqnarray}}
\newcommand{\eea}{\end{eqnarray}}
\newcommand{\baa}{\begin{array}}
\newcommand{\eaa}{\end{array}}

\def\eq#1{{Eq.~(\ref{#1})}}

\def\fig#1{{Fig.~\ref{#1}}}
\newcommand{\intl}{\int\limits}

\newcommand{\bas}{\bar{\alpha}_S}

\newcommand{\nn}{\nonumber}

\newcommand{\h}{\frac{1}{2}}

\newcommand{\x}{\vec{x}}
\newcommand{\vb}{\vec{b}}
\newcommand{\z}{\vec{z}}

\newcommand{\Lb}{\left(}
\newcommand{\Rb}{\right)}

\newcommand{\pp}{\partial}

\renewcommand{\vec}[1]{\boldsymbol{#1}}

\newcommand{\ga}{\gamma}
\newcommand{\dY}{\delta \tilde{Y}}
\newcommand{\dy}{\delta Y_0}

\numberwithin{equation}{section}

\begin{document}
\title{Modified homotopy approach for diffractive production in the saturation region}
\author{Carlos Contreras}
\email{carlos.contreras@usm.cl}
\affiliation{Departamento de F\'isica, Universidad T\'ecnica Federico Santa Mar\'ia,  Avda. Espa\~na 1680, Casilla 110-V, Valpara\'iso, Chile}
\author{Jos\'e Garrido}
\email{jose.garridom@usmcl.onmicrosoft.com}
\affiliation{Departamento de F\'isica, Universidad T\'ecnica Federico Santa Mar\'ia,  Avda. Espa\~na 1680, Casilla 110-V, Valpara\'iso, Chile}
\author{ Eugene ~ Levin}
\email{leving@tauex.tau.ac.il}
\affiliation{Department of Particle Physics, School of Physics and Astronomy,
Raymond and Beverly Sackler
 Faculty of Exact Science, Tel Aviv University, Tel Aviv, 69978, Israel}
\author{Rodrigo Meneses}
\email{rodrigo.meneses@uv.cl}
\affiliation{Escuela de Ingenier\'\i a Civil, Facultad de Ingenier\'\i a, Universidad de Valpara\'\i so,Avda  General Cruz 222 , Valpara\'\i so, Chile}

\date{\today}


\begin{abstract}
In this paper we continue to develop  the homotopy method for solving of the non linear evolution equation for the diffractive production in deep inelastic scattering(DIS). We introduce  part of the nonlinear corrections as a first step of this approach. This simplified nonlinear evolution equation is solved analytically taking into account the initial and boundary conditions for the process. At the second step of our approach we demonstrated that 
the perturbative procedure can be used  for the remaining parts of the non-linear corrections. It turns out that these corrections are small and can be estimated in the regular iterative procedure.
 \end{abstract}
\maketitle

\vspace{-0.5cm}
\tableofcontents






\section{Introduction}

In this paper we continue our search for the regular iteration procedure of solving non-linear equations that in  QCD governs dynamics in the saturation region. In our previous paper\cite{CLMNEW} we developed the homotopy solution \cite{HE1,HE2} for the Balitsky-Kovchegov (BK) equation\cite{BK} that give the dipole scattering amplitude. It has been shown that this approach allows us to collect all essential contribution into the linear equation which can be solved analytically,  and to  propose the iteration  procedure, which is being partly numerical, leads to small corrections. In this paper our main goal   is  to   expand this approach to the processes of diffraction production in deep inelastic scattering (DIS).

The non-linear evolution equation that describes these processes  have been derived in Ref.\cite{KOLE} (see also \cite{HWS,HIMST,KLW,KLP}.The close attention of the high energy community to these processes \cite{KOLE,HWS,HIMST,KLW,KLP,LEWU,GBKW,GOLEDD,SATMOD0,KOML,MUSCH,MASC,MAR,KLMV,LELUDD,LELUDD1,KOLEB,CLMP,CLMP1,MM1,MM2,MM3} shows the need  to find effective procedure for solving the equations.
In derivation of the non-linear equations two approaches  are used: the unitarity constraints for the dipoles scattering amplitude and the AGK\cite{AGK} cutting rules. In 
Ref.\cite{KLP}  has been shown that this non-linear evolution has the same structure for all $\sigma_n$  in DIS, where $\sigma_n $ is the cross section of the produced $n$-dipoles in the final state. Therefore, we view this paper as a starting point for finding $\sigma_n$ for DIS.
 Diffractively we produces the system of dipoles with low multiplicities and the large rapidity gap between them and recoiled target. The total cross section of diffractive production can be viewed as $\sigma_0$: production of particles with multiplicities lower that the average one.
 
 \begin{figure}
 	\begin{center}
 	\leavevmode
 		\includegraphics[width=8cm]{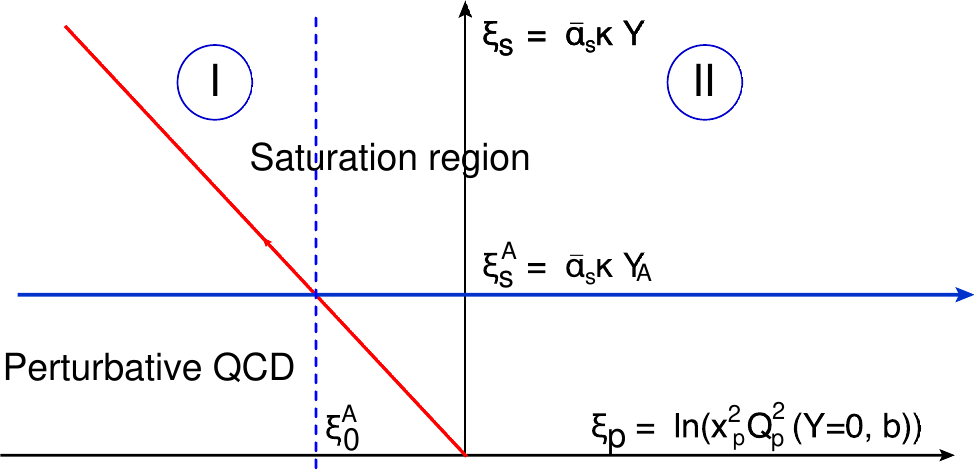}
 	\end{center}
 	\caption{Saturation region of QCD for the elastic scattering amplitude. The critical line (z=0) is shown in red. The initial condition for scattering with the dilute system of partons (with proton) is given at $\xi_s = 0$. For heavy nuclei the initial conditions are placed at $Y_A = (1/3)\ln \,A \gg\,1$, where $A$ is the number of nucleon in a nucleus. The line, where they are given, is shown in blue.}
 	\label{sat}
 \end{figure}
 
 
 The homotopy method we can use for the general equation:
 \beq \label{HOM1}
\mathscr{L}[u] +  \mathscr{N_{L}}[u]=0
\eeq 
 where the linear part $\mathscr{L}[u] $ is a differential  or integral-differential operator, but non-linear part $
 \mathscr{N_{L}}[u] $ has an arbitrary form. As a solution, we introduce  the following  equation for the homotopy function $ {\mathscr H}\Lb p,u\Rb$:
 \beq \label{HOM2}
 {\mathscr H}\Lb p,u\Rb\,\,=\,\,\mathscr{L}[u_p] \,+ \,  p\, \mathscr{N_{L}}[u_p] \,\,=\,\,0
 \eeq
 
 Solving \eq{HOM2} we reconstruct the function
 \beq \label{HOM3}
 u_p\Lb Y,  \x_{10},  \vb\Rb\,\,=\,\, u_0\Lb Y,  \x_{10},  \vb\Rb\,\,+\,\,p\, u_1\Lb Y,  \x_{10},  \vb\Rb \,+\,p^2\, u_2\Lb Y,  \x_{10},  \vb\Rb \,\,+\,\,\dots
 \eeq
 with $\mathscr{L}[u_0] = 0$. \eq{HOM3}  gives  the solution to the non-linear equation at $ p = 1$.  The hope is that several  terms in series of \eq{HOM3} will give a good  approximation in the solution of the non-linear  equation. 
  
 The linear equation is obvious in the perturbative QCD region (see \fig{sat}, where it is the BFKL equation \cite{BFKL,LIP,LIREV}. However,  in Ref.\cite{CLMNEW} we  showed that inside the saturation region (see \fig{sat})  we can find the linearized  equation based on the approach of Ref.\cite{LETU}. In this paper we include in $\mathscr{L}[u_p] $ part of the non-linear corrections which can be treated analytically. 
We  demonstrate that non-linear terms which include the remains of the non-linear corrections, lead to small contributions and can be treated in perturbation approach. 
 
 The detailed description of our approach is included in  the next section. Here we wish 
 to discuss the kinematic region where we are looking for the solution and general assumptions that we make in fixing initial and boundary conditions. In \fig{sat}  the main kinematic regions are shown for the scattering amplitude of a dipole ($x_{10}$)   with  a nucleus target in the plot with $\xi_s$ and $\xi$ axes, where
 $\xi \,\,=\,\,\ln\Lb x^2_{10}\,Q^2_s\Lb Y=Y_A, b \Rb\Rb$ and $ \xi_s = \ln\Lb Q^2_s(Y)/Q^2_s\Lb Y=Y_A, b\Rb\Rb$
 
 The saturation moment  $Q^2_s(Y)$ is equal to
 
 \beq \label{QS}
 Q^2_s\Lb Y, b\Rb\,\,=\,\,Q^2_s\Lb Y=Y_A, b\Rb \,e^{\bas\,\kappa \Lb Y - Y_A\Rb}\
 \eeq 
where $\kappa$ are determined by the following equations\footnote{$\chi\Lb \gamma\Rb$ is the BFKL kernel\cite{BFKL} in anomalous dimension ($\gamma$) representation.  $\psi(\ga)=\Gamma'(\ga)/\Gamma(\ga)$.
}:
 \beq \label{GACR}
\kappa \,\,\equiv\,\, \frac{\chi\Lb \gamma_{cr}\Rb}{1 - \gamma_{cr}}\,\,=\,\, - \frac{d \chi\Lb \gamma_{cr}\Rb}{d \gamma_{cr}}~~~\,\,\,\mbox{and}\,\,\,~~~\chi\Lb \gamma\Rb\,=\,\,2\,\psi\Lb 1 \Rb\,-\,\psi\Lb \gamma\Rb\,-\,\psi\Lb 1 - \gamma\Rb
 \eeq 
   
   In \fig{sat} one can see that for $z\, <\,0$  we have
 the perturbative QCD region where the non-linear corrections are small and we can safely use the BFKL linear equation for the scattering amplitude. $z$ is defined as follows
   \beq \label{ZDY}
z\,\,=\,\,\ln\Lb r^2 \,Q^2_s\Lb \dY, b\Rb\Rb \,\,=\,\,\,\,\,\bas\,\kappa \,\Lb Y\,-\,Y_A\Rb\,\,+\,\,\xi\,\,\,=\,\,\xi_s\,\,+\,\,\xi;~~~~~~~~~~\dY\,\,=\,\,Y \,\,-\,\,Y_A;
\eeq 
with $\xi \,\,=\,\,\ln\Lb Q^2_s\Lb \dY=0\Rb \,r^2\Rb$.

 In \eq{ZDY} and \eq{QS} $Y_A$ denotes $ Y_A = \ln A^{1/3}$, where $A$ is the number of nucleons in a nucleus.

For $z>0$ the non-linear corrections become essential and we enter the saturation region. For the scattering with nuclei, which we consider in this paper, the saturation region can be divided in two parts. For $\xi <\,\xi^A_0$ the amplitude has the geometric scaling behaviour \cite{BALE,GS} and it depends only on one variable $z$. For $\xi\, >\,\xi^A_0$ 
this geometric scaling behaviour is broken. 

\begin{figure}[ht]
    \centering
  \leavevmode
      \includegraphics[width=10cm]{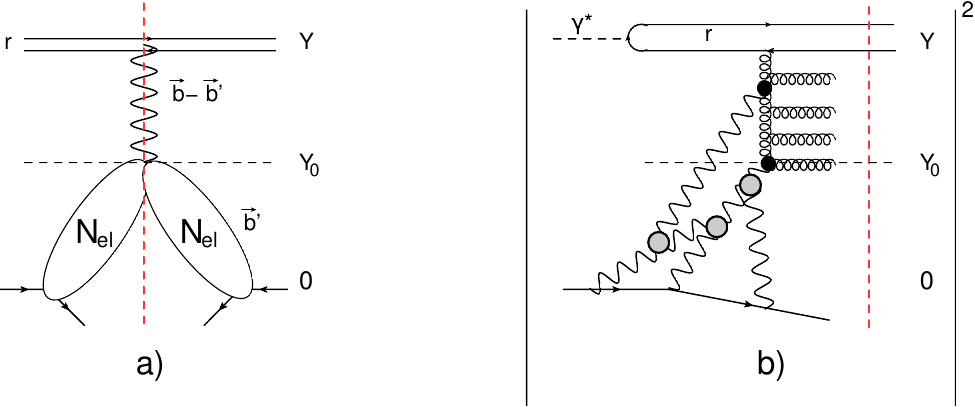}  
      \caption{The graphic representation of the processes of
 diffraction production. \fig{gen}-a shows the first Mueller diagram~\cite{MUDI} for the diffractive production in the scattering of one dipole with size $r$ and rapidity $Y$. In this diagram we indicate the impact parameters $\vec{b}'$ and $\vec{b} - \vec{b}'$ which we used in the text. The wavy lines denote the BFKL Pomeron. The vertical dashed line shows that all gluons of this Pomeron are produced.  \fig{gen}-b shows the general structure of the diagrams that has been taken into account in the equation of Ref.~\cite{KOLE}. In this figure we clarify the notation for the rapidity gap $Y_0$ and the rapidity region $\delta Y = Y - Y_0$ which is filled by produced gluons.
}
\label{gen}
   \end{figure}

For diffractive production we have more complex kinematics (see \fig{gen} for the diffraction production in the region of small mass). From this figure one can see that we have several different kinematic regions for $N^D$ where 
\beq \label{EQ2}
 \sigma_{\rm dipole}^{diff}(r_{\perp},Y, Y_0)
\,\,=\,\,\,\,\int\,d^2 b\,N^D(r_{\perp},Y, Y_0;\vec{b})\,, 
\eeq
is the cross section of diffractive production with the rapidity gap larger than$Y_0$ )(see \fig{gen}).

This process can be characterized by two saturation momentums: $Q_s\Lb Y_0\Rb$ and $Q_s\Lb Y - Y_0\Rb$.  For $ r_{\perp} \,Q_s\Lb Y_0\Rb \,<\,1$  and $ r_{\perp} \,Q_s\Lb Y - Y_0 \Rb\,<\,1$ we can replace $N_{el}$ in \fig{gen}-b by the BFKL Pomeron and the diffraction  cross section  can be calculated   using triple Pomeron diagram.
For $ r_{\perp} \,Q_s\Lb Y_0 \Rb\,>\,1$  and $ r_{\perp} \,Q_s\Lb Y - Y_0 \Rb\,<\,1$   we have the situation which is shown in \fig{gen}. The elastic amplitude is in the saturation region and the production of gluons can be computed using the BFKL Pomeron exchange.
 Finally, for $ r_{\perp} \,Q_s\Lb Y_0\Rb \,>\,1$  and $ r_{\perp} \,Q_s\Lb Y - Y_0\Rb \,>\,1$ 
the large mass is produced and $N^D$ is the kinematic region where non-linear corrections  for gluon production are essential. The main goal of this paper is to find reliable procedure to calculate $N^D$ in this kinematic region. 


 We have to specify the region of $r_{\perp}$ which we are dealing with in the saturation region $ r^2_{\perp}\,Q^2_s\Lb Y,b\Rb\,>\,1$. As has been noted in 
 Refs.\cite{GOST,BEST}  actually for very large $r_{\perp} $ the non-linear corrections are not important and we have to solve linear BFKL equation. This feature can be seen  directly
 from the eigenfunction of this equation.  Indeed, the eigenfunction
 (the scattering amplitude of two dipoles with sizes $r _{\perp}\equiv  x_{10}$ and $R$) has the
 following form \cite{LIP}
\beq \label{EIGENF}
\phi_\gamma\Lb \vec{r}_{\perp} , \vec{R}, \vec{b}\Rb\,\,\,=\,\,\,\Lb \frac{
 r^2\,R^2}{\Lb \vec{b}  + \h(\vec{r}_{\perp} - \vec{R})\Rb^2\,\Lb \vec{b} 
 -  \h(\vec{r}_{\perp} - \vec{R})\Rb^2}\Rb^\gamma  \,\xrightarrow{ b \gg\,r, R} \Lb\frac{R^2 r^2}{b^4}\Rb^\gamma~~\mbox{with}\,\,0 \,<\,Re\gamma\,<\,1
 \eeq 

One can see that for $r_{\perp}=x_{10} \,>\,min[R, b]$,  $\phi_\gamma$ starts to be smaller and the non-linear term in the BK equation 
could be neglected.  In this paper we consider the interaction with the  nucleus  and the  scattering amplitude of the dipole with a nucleus  for the exchange of the BFKL Pomeron   has the following form:
\beq \label{EIGENF1}
A_{d A} \Lb  \vec{r}_{\perp} , \vec{R}_N, \vec{b}\Rb\,=\,\int\phi_\gamma\Lb \vec{r}_{\perp} , \vec{R}_N, \vec{c}\Rb\,\,S_A\Lb \vec{b} - \vec{c}\Rb d^2 c\eeq
where $R_N$ is the size of a nucleon, $\vec{c}$ is the impact parameter for dipole-nucleon amplitude and $\vec{b} - \vec{c}$ is the position of the nucleon with respect to the center  of the nucleus. $S_A\Lb \vec{b} - \vec{c}\Rb$ is the number of  nucleons in a nucleus.  Since $c \,\ll \,b$   we can integrate over $\vec{c}$ replacing $\vec{b} - \vec{c} $ by $\vec{b}$ and obtain  the expression:
\bea \label{EIGENF2}
A_{d A} \Lb  \vec{r}_{\perp} , \vec{R}_N, \vec{b}\Rb  &=&\int \phi_\gamma\Lb \vec{r} , \vec{R}_N, \vec{c}\Rb\,\,S_A\Lb \vec{b} - \vec{c}\Rb d^2 c = \int d^2 c \, \phi_\gamma\Lb \vec{r}_{\perp} , \vec{R}_N, \vec{c}\Rb\,\,S_A\Lb \vec{b}\Rb \nn\\
&=&  \Lb \frac{r^2_{\perp}}{R^2_N}\Rb^\gamma \!\!R^2_N S_A\Lb \vec{b}\Rb   =   \Lb r^2_{\perp}\,Q^2_s\Lb Y=Y_A, b\Rb \Rb^\gamma\,\eea
Therefore, in this case we can absorb all dependence on the impact parameter in the $b$  dependence of the saturation scale. In \eq{EIGENF2} we implicitly assume that $ r_{\perp} \leq R_N$. 
 The typical process, that we bear in mind, is the deep inelastic scattering(DIS) with a nucleus at $Q^2\, \geq \,1 GeV^2$ and at small values of $x$.


\section{The main equation} 
\begin{figure}[ht]
    \centering
  \leavevmode
      \includegraphics[width=14cm]{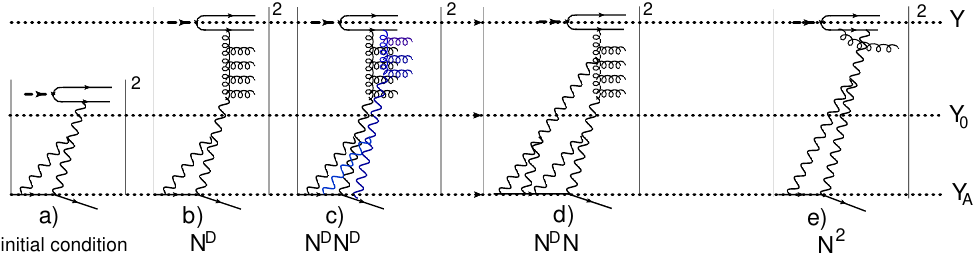}  
      \caption{The graphic representation of the terms of \eq{SDEQ} for
 diffraction production. \fig{gen1}-a shows   the initial conditions for $N^D$. 
 \fig{gen1}-b is the linear  equation for $N^D$, while \fig{gen1}-c describes the non-linear contribution $N^D\,N^D$. \fig{gen1}-d shows the shadowing correction due to enhanced diagrams. The wavy lines denote the BFKL Pomeron while the helix ones describe the gluons.
}
\label{gen1}
   \end{figure}

 The equation for $N^D\Lb \dY, \delta Y_0,r_{10}; b \Rb$  has the following form\cite{KOLE}:
 \bea \label{SDEQ}
&& \frac{\partial N^D\Lb Y, Y_0,r_{10}; b \Rb}{\partial Y_M}\,=\,\\
&&
\,\,\frac{\bas}{2 \pi}\,\int\,d^2 r_2 \,K\Lb r_{10}| r_{12}, r_{02}\Rb \left\{N^D\Lb Y, Y_0,r_{12}; b \Rb\,+\,\begin{scriptsize}\begin{footnotesize}\end{footnotesize}\end{scriptsize}
N^D\Lb Y, Y_0, r_{20}; b \Rb\,-
\,N^D\Lb Y,  Y_0, r_{10}; b \Rb\,\right.\nn  \\
&& \left.+
N^D(Y; Y_0, r_{12}; b)  N^D(Y; Y_0, r_{20}; b)
\,-\,2\,N^D(Y; Y_0, r_{12}; b )\,N(Y;r_{20}; b) -\,2\,
N(Y;r_{12}; b)
N^D(Y; Y_0, r_{20}; b)\right.\nn\\
&&\left.
\,+\, 2
N(Y; r_{12}; b)
N(Y;r_{20}; b)
\right\}\notag
 \eea    
 In this equation $N(Y;r_{01}; b) $ is the imaginary part of the elastic scattering amplitude of the dipole with the size $r_{01}$ and rapidity $Y$at the impact parameter $b$ (see \fig{gen}).  $N^D(Y; Y_0, r_{01}; b) $ is the cross section of the diffractive production with the rapidity gap larger than $Y_0$ for the same scattering process. The  cross section of the diffractive production of particles (gluons) in rapidity region $Y_M = Y - Y_0$ ($\sigma_{diff}$, see \fig{gen1})  can be found from the following equation
 \beq \label{DIFF}
N^D(Y; Y_0, r_{01}; b)\,\,=\,\,N^2_{el}\Lb Y_0\Rb + \intl^{Y}_{Y_0 }d Y' 
\sigma_{diff}\Lb Y',Y_0,b\Rb       \equiv   N^2_{el}\Lb Y_0\Rb + \intl^{Y_M}_{0 }d Y' n^D(Y_M', Y_0, r_{01}; b) \eeq 
 
 From \eq{DIFF} one can see that \eq{SDEQ} gives $n^D$:
\bea \label{DIFF1}
 n^D_{01}\,&=&\,\,\,\frac{\bas}{2 \pi}\,\int\,d^2 r_2 \,K\Lb r_{10}| r_{12}, r_{02}\Rb \Bigg\{
N^D_{12}+N^D_{02}\,-
\,N^D_{01}+
N^D_{12}  N^D_{02}
\,-\,2\,N^D_{12}\,N_{02} -\,2\,
N_{12}
N^D_{02}
\,+\, 2
N_{12}
N_{02}\Bigg\}
 \eea 
where  $n^D_{01}$,  $N_{01}^D $ and $N_{01}$ is the obvious shorthand for $n^D(Y, Y_0, r_{01}; b)$, $N^D(Y, Y_0, r_{01}; b)$ and $ N(Y; r_{01}; b)$ respectively. In \fig{gen1} we pictured each of the contribution in \eq{SDEQ} and \eq{DIFF1}.

For large $\dy = Y_0 -Y_A$ the elastic amplitude is  in the saturation region and close to 1. Replacing $N_{ij} =1$ and neglecting term $N^D N^D$ for $Y  \to Y_0$
(see \fig{gen1}-c) one can see that \eq{DIFF1} reduces to the linear equation

\beq \label{DIFF2}
 \frac{\partial N_{01}^D}{\partial Y_M} =\bas\int \frac{d^2x_{2}}{2\pi}\frac{x_{01}^2}{x_{02}^2x_{12}^2}\Lb -N_{12}^D-N_{02}^D-N_{01}^D\,+2\Rb =
 -\bas\int \frac{d^2x_{2}}{2\pi}\frac{x_{01}^2}{x_{02}^2x_{12}^2}\Lb N_{12}^D+N_{02}^D- N_{01}^D\Rb \,-\,\, 2z_M\,(N_{01}^D - 1)\
 \eeq
 The last term is equal to\cite{LETU}
 \beq \label{DIFF20} 
  -\bas\int \frac{d^2x_{2}}{2\pi}\frac{x_{01}^2}{x_{02}^2x_{12}^2}\Lb  N_{01}^D\,-\,1\Rb =
  -\bas \int^{x^2_{01}}_{1/Q^2_s (Y_M) }  \frac{d x^2_{02}}{x^2_{02}}\Lb  N_{01}^D\,-\,1\Rb  \eeq
  where $Q^2_s\Lb Y_M\Rb $ is the saturation momentum. It is well known that to find the saturation momentum we need to consider  the linear equation, which correspond to \fig{gen1}-b. It can be written as (see \eq{EIGENF} and Ref.\cite{CLMS})
  \beq \label{DIFF21}
  n^D\Lb \fig{gen1}-b\Rb = \intl^{\epsilon + i \infty}_{\epsilon - i \infty}\frac{ d \gamma}{2 \pi i}\exp\Lb \bas \chi\Lb \gamma\Rb Y_M\Rb \phi_\gamma\Lb \vec{r}_{\perp} , \vec{r}^{\prime}, \vec{b}\Rb N^2\Lb \dy, r'\Rb\, \frac{d^2 r^{\prime}  d^2 b }{r'^4}
  \eeq
  Taking this integral using the method of steepest decent (see Refs.\cite{GLR,MUT}) we obtain that $Q^2_s\Lb Y_M\Rb = \frac{1}{r'^2} \exp\Lb \bas \,\kappa Y_M\Rb$.  We need to take integral over $r'$ in \eq{DIFF21} to find the typical values of $r'$. First one can see that integral over $b$ (see \eq{EIGENF})  leads to extra factor $r'^2$ since $r' > r$. Finally,
   the integral over $r'$ has a form:
   \beq \label{DIFF22} 
   \int \frac{d r'^2}{r'^2}  \Lb r'^2\Rb^{- \gamma} N^2\Lb \dy, r'\Rb\,\,\sim \,\,\Lb r'^2_{min} = 1/Q^2_s\Lb \dy\Rb\Rb^{- \gamma}  
   \eeq
   since $N^2\Lb \dy,r'\Rb = 1 $ for $r'\, > \,1/Q_s\Lb \dy\Rb$.
   
   Bearing this in mind integral in \eq{DIFF20} gives $z_M$:
 
 \beq \label{ZM}
 z_M\,\,=\,\,\ln\Lb r^2 \,Q^2_s\Lb Y_M, b\Rb\Rb \,\,=\,\,\,\,\,\bas\,\kappa Y_M\,\,+\,\,\xi_M\,\,\,~~~~~~~~~~Y_M\,\,=\,(\,Y \,\,-\,\,Y_0 ) ~~ \mbox{and}~~\xi_M=\ln\Lb r^2\,Q^2_s\Lb\dy\Rb\Rb;
\eeq 
 Introducing $\xi$ of \eq{ZDY} one can see that $z_M = z$.
 From \eq{DIFF2} we can conclude that $N^D$ decreases  with $Y_M$.

 It turns out that \eq{SDEQ} can be rewritten in this  simple form, 
 introducing a new function:
\begin{equation}\label{AMPLITUD}
	{\mathscr N}(z,\dY, \delta Y_0)\, =\,  2\, N(z,\dY)\, -\, N^D(z, \dY, \delta Y_0)
\end{equation}
where  $N\Lb z, \dY\Rb$ is the solution to the BK equation and $z$ is given by \eq{ZDY}. The new variables $\dY$ and $\delta Y_0$ are defined as $\dY= \bas\Lb Y - Y_A\Rb$ and $\delta Y_0 = \bas \Lb Y_0 - Y_A\Rb$.
This function has  a clear  meaning: the inelastic cross section of all events with the rapidity gap from $Y=0$ to   $Y=Y_0$. 
 
 For function ${\mathscr N}$ \eq{SDEQ} takes the form of the BK equation, viz.:
 \beq \label{SDBK}
 \frac{\partial{\mathscr N}_{01}}{\partial Y}\,=\,\bas\int \frac{d^2\,x_{02}}{2 \pi} \frac{ x^2_{01}}{x^2_{02}\,x^2_{12}}\Big\{ {\mathscr N}_{02} + {\mathscr N}_{12} - {\mathscr N}_{02} {\mathscr N}_{12} - {\mathscr N}_{01}\Big\}
 \eeq 
 The initial conditions for this equation is the following:
 \beq \label{IC}
{\mathscr N}(z \to z_0,\dY = \delta Y_0  , \delta Y_0)\,\,=\,\, 2\, N(z_0,\delta Y_0) \,\,-\,\, \, N^2(z_0,\delta Y_0)
\eeq
\eq{IC} can be seen directly from \fig{gen}.

The initial condition where  $z =z_0$, which corresponds to $\dY =   \delta Y_0$, has the following form \cite{MUT}(see \fig{sat}):
  \beq \label{BC}
  {\mathscr N}(z \to z_0 ,\dY=\delta Y_0, \delta Y_0)\,\,=\,N_{in}\Lb r_\perp, \delta Y_0\Rb
  \eeq
  
  where $N_{in}\Lb r_\perp, \delta Y_0\Rb\,\,\equiv\,\,2\, N(z_0,\delta Y_0) \,\,-\,\, \, N^2(z_0,\delta Y_0)$.

  For large values of $\delta Y_0$ $N_{in}$ tends to\cite{LETU} 
   \beq \label{BC1}
  N_{in}\Lb r'_\perp, \dy\Rb \,\xrightarrow{ z_0 \gg 1} 1 \,-\,C^2 \exp\Lb - \frac{z^2_0}{\kappa}\Rb ~~~~~~~~\mbox{with}~~~ z_0 \,\,=\,\,    
\ln\Lb r'^2 \,Q^2_s\Lb \dy, b\Rb\Rb \,\,=\,\,\,\,\,\kappa \,\dy\,\,+\,\,\xi ;     \eeq
where $C$ is a smooth function of $z_0$. However, \eq{BC1}  only holds  in  region I in \fig{sat}, while in region II we have to use a more general expressions for the elastic scattering amplitudes (see Ref.\cite{CLMNEW}). We will discuss the initial  condition in this region below.

Our strategy of finding this solution looks as follows: first we are going to solve \eq{SDBK} and find $\mathscr{ N}_{01} \,=\, 1 - \Delta^D$, and after that we will return to \eq{SDEQ}.  
\begin{boldmath}
\section{Modified homotopy approach for $\Delta^D$ } 
\end{boldmath}

\begin{boldmath}
\subsection{Generalities }
\end{boldmath}

As previously  mentioned we use the homotopy approach to find the solution to non-linear equation of \eq{SDBK}, suggested in Refs.\cite{HE1,HE2,CLMNEW}.       
     In this paper we modify the homotopy approach by including  the part of the non-linear term into definition of $ \mathscr{L}[u_0] $.

      Following the main ideas of Ref.\cite{LETU} we solve \eq{SDBK} replacing $ \mathscr{N}\Lb z, \delta Y_0\Rb $ by  $ \mathscr{N}\Lb z, \delta Y_0\Rb\,\,=\,\,1\,\,-\,\,\Delta^D\Lb  z, \delta Y_0\Rb$ . For  this function the equation takes the form:
      \beq\label{SDBK1}
 \frac{\partial  \Delta^D_{01}}{\partial Y}\,=\,\bas\int \frac{d^2\,x_{02}}{2 \pi} \frac{ x^2_{01}}{x^2_{02}\,x^2_{12}}\Big\{  \Delta^D_{02} \Delta^D_{12} - \Delta^D_{01}\Big\}
\eeq 
   and the  initial conditions for $\Delta^D$ takes the following form\cite{CLMS}:
    \begin{subequations}    \bea 
  {\rm Region\,~I\,}: &&  \Delta^D_{01}\Lb z \to z_0,\delta Y_0\Rb\,\,=\,\,C^2\exp\Lb - \frac{\Lb z_0  - \tilde{z}\Rb^2}{\kappa} \Rb \label{ICBK1} \\
    {\rm Region~II}: &&\Delta^D_{01}\Lb z \to z_0,\dY \to \dy,\delta Y_0\Rb=  1\,-\,N_{in} \,\,=\,\,
 G^2\Lb \xi\Rb \exp\Lb - \frac{\Lb z_0 - \tilde{z}\Rb^2}{\kappa} \Rb\label{ICBK2} \\ 
\mbox{with} &&G\Lb \xi\Rb\,\,=\exp\Lb
 \frac{\Lb \xi - \tilde{z}\Rb^2}{2\,\kappa} \,\,-\,\,\frac{1}{4}e^{\xi}\Rb
 \label{RIIEL1} \eea
  \end{subequations} 
  
    where $\xi$ is defined as $\xi =\ln\Lb Q^2_s\Lb Y_A\Rb r^2\Rb$  and $\tilde{z} = 2 (\ln2 + \psi(1))\,-\,\zeta$ .
     
   We suggest  to simplify the non-linear term replacing  it with
    \beq \label{SDBK12}
\,\bas\int \frac{d^2\,x_{02}}{2 \pi} \frac{ x^2_{01}}{x^2_{02}\,x^2_{12}} \Delta^D_{02} \Delta^D_{12} \,\,\to\,\,\Delta^D_{01} \intl^z_0 d z'  \Delta^D_{02}\,\,=\,\,  \Delta^D_{01} \Bigg( \zeta - \intl^\infty_z d z'  \Delta^D_{02}\Bigg)~~\mbox{with}~~\zeta = \intl^\infty_0d z'  \Delta^D_{02}\eeq    
with $z'$ which is determined by \eq{ZDY} by replacing $\xi $ by $ \ln\Lb Q^2_s\Lb \dY=0\Rb x^2_{02}\Rb$. This contribution stems from the region $x_{02}\,\ll\,x_{01}$ (see Ref.\cite{LETU}) and 
we believe that   \eq{SDBK12} is a good approximation for the integral. W]e can treat the corrections to this equation  in a perturbation approach.
       
       We modify the homotopy approach :
      taking  
      \beq \label{LBK}
     \mathscr{L}\Lb \Delta^D_0\Rb\,\,=\,\,\Lb \frac{\pp}{\pp \dY} +\,\,z\,\,-\,\,\zeta\Rb\,\Delta^D_0+\,\,\Delta^D_0\Lb z, \dY, z_0\Rb \intl^\infty_z  d z'  \Delta^D_{0} \Lb z',\dY, z_0\Rb \,
     \eeq
      we find the solution to the equation:
      \beq \label{SDBK2}
        \mathscr{L}\Lb \Delta^D_0\Rb\,\,=\,\,0;~~~~
             \Lb \frac{\pp}{\pp \dY} +\,\,z\,\,-\,\,\zeta\Rb\,\Delta^D_0+\,\,\Delta^D_0\Lb z, \dY, z_0\Rb \intl^\infty_z  d z'  \Delta^D_{0} \Lb z', \dY, \z_0\Rb\,=\,0;        
  \eeq

  It turns out that searching for the function $\Delta^{(0)}\Lb z ,\dY\Rb$ it is more transparent to
 return to \eq{SDBK}, which can be rewritten in the region where  $x_{02}\,\ll\,x_{01}$ or  $x_{12}\,\ll\,x_{01}$ in the form (see Ref.\cite{LETU} and section II-C-2 of Ref.\cite{CLMS}):
\beq\label{GSS1}
\kappa \frac{ d \mathscr{N}_{01} \Lb z, \xi_s \Rb}{d \xi_s}= \Lb1 - \mathscr{N}_{01}\Lb \xi, \xi_s \Rb\Rb\intl^\xi_{- \xi_s} d \xi'	\mathscr{N}_{02}\Lb \xi', \xi_s\Rb	;~~~\mbox{with}~~~\xi_s = \kappa \,\dY
\eeq

Introducing $\Delta^{(0)}\Lb z,  \xi_s\Rb =  1 - \mathscr{N}_{01}\Lb z, \xi_s\Rb = \exp\Lb - \Omega^{(0)}\Lb z, \xi_s\Rb\Rb$ we obtain:
\beq\label{GSS2}
	\kappa\dfrac{d \Omega^{(0)}\Lb z, \xi_s\Rb}{d \xi_s} = 
	\intl^z_{0} d z' \Lb 1 - e^{ - \Omega^{(0)}\Lb z', \xi_s\Rb}\Rb; ~~~
	\kappa\dfrac{\pp^2\Omega^{(0)}\Lb z, \xi_s\Rb}{\pp \xi_s \,\pp z} 	\,\,=\,\,1 -  e^{ - \Omega^{(0)}\Lb \z,\xi_s\Rb}
	 \eeq~~

\begin{boldmath}
\subsection{Solutions to the master equation }
\end{boldmath}
\begin{boldmath}
\subsubsection{Geometric  scaling solution to the master equation}
\end{boldmath}

Searching for the function $\Delta^{(0)}\Lb z,z_0 \Rb$ we can rewrite \eq{GSS2} in the form\footnote{It is worthwhile mentioning that this solution 
give the violation of the geometric scaling behaviour of the scattering amplitude due to dependence of $\Omega$ on $z_0$ and $z$.}:
\beq \label{GSSS21}
\kappa\dfrac{d^2\Omega^{(0)}\Lb z, z_0\Rb}{d z^2} 	\,\,=\,\,1 -  e^{ - \Omega^{(0)}\Lb z, z_0\Rb}
\eeq

Assuming that $\dfrac{d \Omega^{(0)}\Lb z\Rb}{d z}  =p\Lb \Omega^{(0)}\Rb$	
	\eq{GSS2} can be rewritten as

\beq\label{GSS3}
\h \kappa\, \frac{d p^2}{d \Omega^{(0)}} \,=\,1 -  e^{ - \Omega^{(0)}\Lb z\Rb}\eeq 
with the solution:

\beq\label{GSS4}
p\, =\dfrac{ d  \Omega^{(0)}}{d  z}\,\,= \,\sqrt{ \frac{2}{\kappa}\Lb \Omega^{(0)}\,+\,\exp\Lb - \Omega^{(0)}\Rb - 1\Rb + C_1}
\eeq
where $C_1$ is a constant.  Integrating \eq{GSS4} we have
\beq\label{GSS5}
	\intl^{\Omega^{(0)}}_{\Omega^{(0)}_0}\frac{ d \Omega'}{\sqrt{ \Omega' + \exp\Lb - \Omega'\Rb -  \Omega^{(0)}_0}}\,\,=\,\sqrt{\frac{2}{\kappa}}\,\Lb z + C_2\Rb
	\eeq
Constant  $C_2$  as well as $\Omega^{(0)}_0$   can be found from 	the solution of the linear equation. Assuming that $\Omega^{(0)}_0$ is large we can neglect the contribution of $\exp\Lb - \Omega'\Rb $ in \eq{GSS5}. Solution in this kinematics we know (see Refs.\cite{CLMP,CLMP1}):
\beq\label{GSS50cc}
\Omega^{(0)}(z, z_0) = \Lb z - \tilde{z} \Rb^2/(2\,\kappa)  + \Lb z_0 - \tilde{z} \Rb^2/(2\,\kappa)-\,2\ln\Lb C \Rb.
\eeq	
One can see that it comes from \eq{GSS5} for
\beq\label{GSS50}
\Omega^{(0)}_0=  \Lb z_0  - \tilde{z}\Rb^2/(2\,\kappa) \,-\,2\ln\Lb C \Rb;~~~ ~~C_2= - \tilde{z}.
\eeq		
Finally,  the solution with geometric scaling behaviour  can be found from the following implicit equation:
\beq\label{GSS8}
\mathscr{U}\Lb\Omega^{(0,I)}, \Omega^{(0)}_0 \Rb\,\,=\,\,\sqrt{\frac{2}{\kappa}}\Lb z - \tilde{z}\Rb
\eeq
where
\beq\label{GSS81}
\mathscr{U}\Lb\Omega^{(0,I)}, \Omega^{(0)}_0 = a\Rb\,\,\,=\intl^{\Omega^{(0)}}_{a}\!\!\!\frac{ d\, \Omega'}{ \sqrt{\Omega' + \exp\Lb - \Omega'\Rb  - a}}
\eeq
It worthwhile mentioning that this solution is the traveling wave solution to \eq{GSS3}(see {\bf 3.4.1.1} of Ref.\cite{MATH2} or  {\bf 2.9.1.1} of Ref.\cite{MATH1}).

~

~

\begin{boldmath}
\subsubsection{General solution to the master equation}
\end{boldmath}

\eq{GSS2} can be rewritten as
 \beq \label{GSG1}
\dfrac{\pp^2\Omega^{(0)}\Lb z, \xi_s\Rb}{\pp z^2} \,\,-\,\,\dfrac{\pp^2\Omega^{(0)}\Lb z, \xi_s\Rb}{\pp t^2}	\,\,=\,\,\frac{1}{\kappa}\Lb 1 -  e^{ - \Omega^{(0)}\Lb z',\xi_s\Rb}\Rb
\eeq
where $z = \xi_s + \xi$ and $t = \xi_s - \xi$. 

This equation has 
 the traveling wave solution (see {\bf 3.4.1.1} of Ref.\cite{MATH2}:
 \beq \label{GSG2}
 \displaystyle{ \intl^{\Omega^{(0)}\Lb z, \xi_s\Rb}_{ \Omega^{(0)}_0\Lb z_0, \xi_{0,s}\Rb} \frac{ d \Omega'}{\sqrt{C_1 + \frac{2}{\kappa ( \mu^2 - \nu^2)}\Lb \Omega' + \exp\Lb - \Omega'\Rb\Rb}}= \mu\,z\,\,+\,\,\nu \,t\,\,+\,\,C_2}
 \eeq
  \eq{GSG2} can be rewritten  as follows\footnote{ In \eq{GSG3} the parameters  $\mu$ , $\nu$,$C_1$ and $C_2$ are redefined in comparison with \eq{GSG2} but we keep the same notations for simplicity.}
  but of  to satisfy the initial condition of \eq{ICBK2}:
  \beq \label{GSG3}
 \displaystyle{ \intl^{\Omega^{(0)}\Lb z, \xi_s\Rb}_{ \Omega_0}\!\!\!\!\!\!\!\! \frac{ d \Omega'}{\sqrt{- \Omega_0 \,+\, \Omega' + \exp\Lb - \Omega'\Rb}}= \sqrt{ \frac{2}{\kappa}}\Lb \Lb 1 + \nu\Rb\,z \,+\,\nu \,t \,-\, \tilde{z}- 2 \nu \,\xi_{0,s}\Rb}
 \eeq 
  where $ \xi_{0,s} = \bas\,\kappa\,\dy$ and
    \beq \label{GSG4}  
 \Omega_0\,\,=\,\,\frac{\Lb z_0 - \tilde{z}\Rb^2}{2\,\kappa} \,-\,\frac{\Lb z_0 -\xi_{0,s} - \tilde{z}\Rb^2}{\,\kappa}  \,+\,\h \exp\Lb  z_0 - \xi_{0,s} \Rb
 \eeq 
Indeed, for $ z \to z_0$ and $ \dY \to \dy$    we can neglect $\exp\Lb - \Omega' \Rb$ in   \eq{GSG3} and this equation gives
  \bea \label{GSG40}     
         2 \sqrt{ \Omega\Lb z, \xi_s\Rb - \Omega_0} &= &\sqrt{\frac{2}{\kappa}}    \Lb\Lb 1 + \nu\Rb\,z \,+\,\nu \,t \,-\, \tilde{z}- 2 \nu \,\xi_{0,s}\Rb;\nn\\   
   ~\mbox{or}~~ 
     \Omega\Lb z, \xi_s\Rb  & =& \frac{1}{2 \kappa} \Lb   \Lb 1 + \nu\Rb\,z \,+\,\nu \,t \,-\, \tilde{z}- 2 \nu \,\xi_{0,s}\Rb^2 + \Omega_0;  \nn\\              
   \mbox{For}\,\,  z = z_0, \xi_s = \xi_{0,s} ~~     \Omega\Lb z_0, \xi_{0,s}\Rb &=& \frac{\Lb z_0 - \tilde{z}\Rb^2}{2 \kappa} + \Omega_0 =\frac{\Lb z_0 - \tilde{z}\Rb^2}{ \kappa} \,+\,\h \exp\Lb  z_0 - \xi_{0,s} \Rb;          
\eea
 one can see that $\nu$ is not  determined by the initial condition and it has to be found from the boundary conditions on the line $\xi = \xi^A_0$.

\newpage

\begin{boldmath}
\subsection{Initial and boundary conditions}
\end{boldmath}
 ~
  
 
  \begin{boldmath}
\subsubsection{Region I}
\end{boldmath}


We have found $\Omega^{(0,I)}\Lb z \Rb $ (see\eq{GSS5} ) which satisfies the boundary conditions of \eq{ICBK1}. 
 The implicit relation of \eq{GSS5}  for $\Omega^{(0,I)}$   can be resolved for large $\Omega^{(0,I)}$. Indeed, expanding
 \bea\label{GSS10} 
 \frac{ 1}{ \sqrt{\Omega' + \exp\Lb - \Omega'\Rb  - \Omega\Lb z_0, \xi_{0,s}\Rb}} & =& \frac{ 1}{\sqrt{ \Omega' + \exp\Lb - \Omega'\Rb  - a} }\\
    &  =& \frac{1}{\sqrt{\Omega' -a}}\Lb 1+\sum^\infty_{k=1} \frac{(2k -1)!!}{k!}\Lb - \frac{e^{ - \Omega'} }{2\Lb\Omega'-a\Rb} \Rb^k\Rb\nn
    \eea
    
\begin{figure}
 	\begin{center}
	\begin{tabular}{ c c}
 	\leavevmode
 		\includegraphics[width=7.5cm]{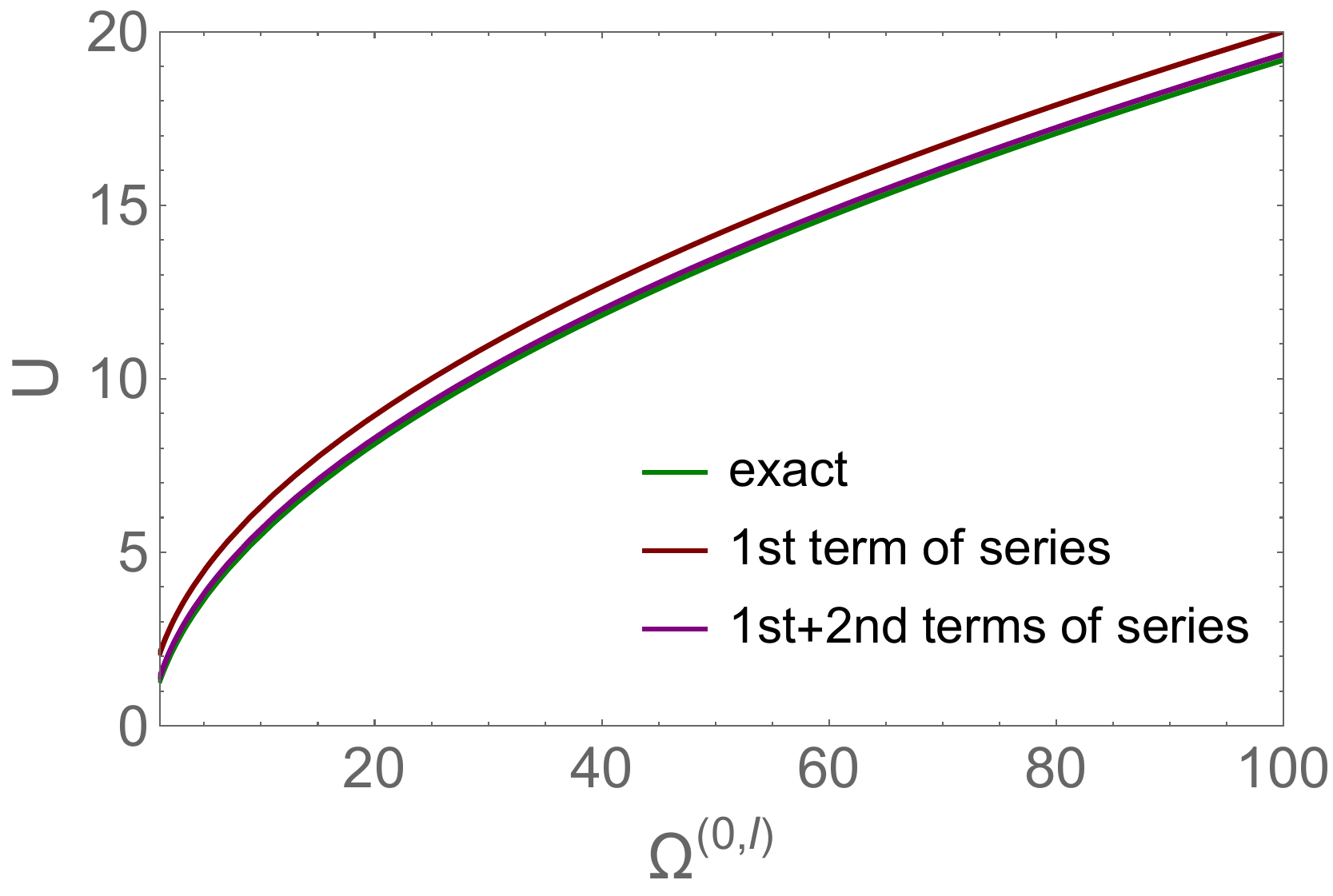}&\includegraphics[width=7.5cm]{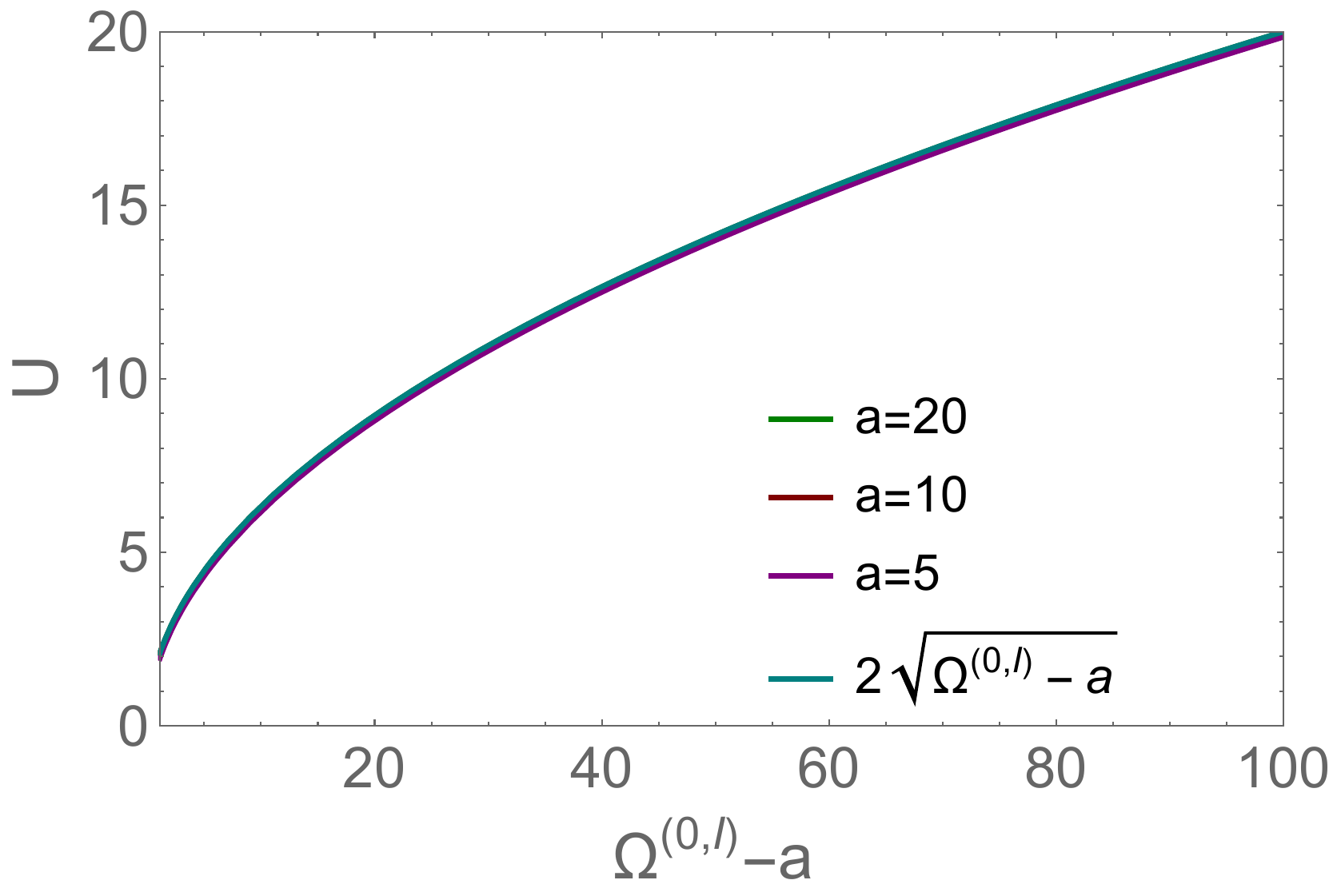} \\
		\fig{u1}-a & \fig{u1}-b\\
		\end{tabular}
			\end{center}
 	\caption{\fig{u1}-a: Function  $\mathscr{U}\Lb\Omega^{(0,I)}\Rb$ versus
	 $\Omega^{(0,I)}$. \fig{u1}-b: Function  $\mathscr{U}\Lb\Omega^{(0,I)}\Rb$ versus
	 $\Omega^{(0,I)}$ at different values of $a$}  
 	\label{u1}
 \end{figure}

We rewrite $\mathscr{U}\Lb\Omega^{(0,I)}\Rb$ as
    \beq \label{GSS101}
    \mathscr{U}\Lb\Omega^{(0,I)}\Rb    =\intl^{\Omega^{(0,I)}}_{a}\frac{ d \Omega'}{\sqrt{ \Omega' + \exp\Lb - \Omega'\Rb -a }} \eeq   
   Plugging \eq{GSS10} in \eq{GSS101} for function  $  \mathscr{U}_1\Lb\Omega^{(0,I)}\Rb$ we obtain:
  \beq\label{GSS11} 
  \mathscr{U}\Lb\Omega^{(0,I)}\Rb=2 \sqrt{\Omega^{(0,I)} - a} +  \sum^\infty_{k=1} \frac{(-1)^{k}k^{k-\h}(2k -1)!!}{2^k\,k!} e^{-a\, k} \Lb \frac{(-1)^k 2^k \sqrt{\pi}}{(2 k -1)!!} - \Gamma\Lb \h -k,(k (-a+\Omega^{(0,I)})\Rb\Rb   \eeq
 \eq{GSS11} gives the series representation of function $  \mathscr{U}\Lb\Omega^{(0,I)}\Rb$. \fig{u1} shows that this series  approximate $ \mathscr{U}\Lb\Omega^{(0,I)}\Rb$ quite well even if we restrict ourselves by first two terms. Actually, one can see from \fig{u1} that at large $a$ the first term in \eq{GSS11} approximates the exact solution quite well.
  
  For large values of $ \Omega^{(0,I)}$ the  \eq{GSS10}   
 takes the form:
   \beq \label{GSS12}
     2 \sqrt{\Omega^{(0,I)} - a} +  \sum^\infty_{k=1} \frac{(-1)^{k-1}(2k -1)!!}{2^k k\,k!}\Lb \frac{1}{\Omega^{(0,I)}-a}\Rb^{k +\h} \exp\Lb - k\,  \Omega^{(0,I)}\Rb\,=\,\,\,\sqrt{\frac{2}{\kappa}}\Lb z - \tilde{z} \Rb
     \eeq
   For large $z$ \eq{GSS12} leads to
   \beq \label{GSS13}
   \Omega^{(0,I)}\Lb z \Rb\,-\,a\,\,=\,\,\frac{ \Lb z\,-\, \tilde{z}\Rb^2}{2\,\kappa} -  \sum^\infty_{k=1} \frac{(-1)^{k-1}(2k -1)!!}{2^k k\,k!}\Lb \frac{1}{\frac{ \Lb z  - \tilde{z}\Rb^2}{2\,\kappa}}\Rb^{k +\h} \exp\Lb - k\,\frac{ \Lb z  - \tilde{z}\Rb^2}{2\,\kappa}\Rb 
   \eeq
   \eq{GSS13} can be rewritten for $\Delta^{(0,I)}_0$ as follows
    \beq \label{GSS14}  
    \Delta^{(0,I)}_0\Lb z \Rb\,\,=\,\,\Delta_{LT}\Lb z\Rb\exp\Lb - a  + \sum^\infty_{k=1} \frac{(-1)^{k-1}(2k -1)!!}{2^k k\,k!}\Lb \frac{2 \kappa}{ \Lb z - \tilde{z}\Rb^2}\Rb^{k +\h} \Delta^k_{LT}\Lb z \Rb  \Rb
    \eeq 
     where $ \Delta_{LT}\Lb z\Rb    \,=\,\exp\Lb - \frac{\Lb z \,-\,\tilde{z}\Rb^2}{2\,\kappa}\Rb$.
       
  For finding $   \Omega^{(0,I)}\Lb z\Rb$  in a general case we have to find the inverse function for $\mathscr{U}_1$. Indeed,
   \beq \label{GSS17} 
  \mathscr{U}\Lb\Omega^{(0,I)}\Rb = \sqrt{\frac{2}{\kappa}}  \Lb z  - \tilde{z}\Rb  \eeq
  where $\Omega^{(0,I)}_0$ is given by \eq{GSS50}.    
  
  
 \begin{figure}
 	\begin{center}
 	\leavevmode
 		\includegraphics[width=14cm]{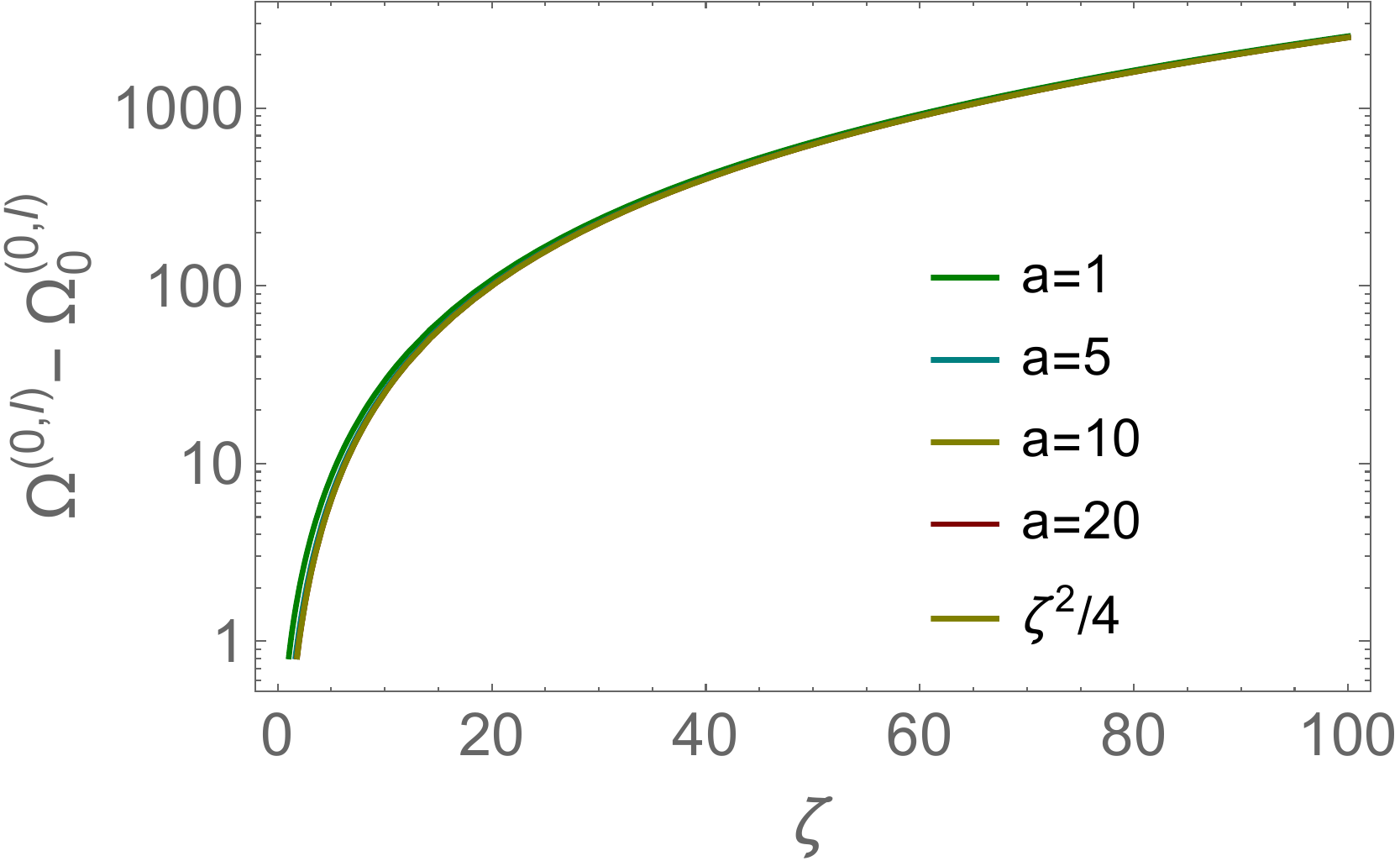}
 	\end{center}
 	\caption{ Function  $\Omega^{(0,I)}$   from \eq{GSS17} versus
	 $\zeta= (z - a )/\sqrt{\h\kappa}$ for different values of $a$.}  
 	\label{om}
 \end{figure}

 
 In \fig{om} we plot $\Omega^{(0,I)}$  versus $  \zeta=\sqrt{ \frac{2}{\kappa}}\Lb z  -\tilde{z}\Rb -a $,  where $a$ is given by \eq{GSS101}. One can see that the first iteration of \eq{GSS14} describes the exact evaluation quite well.
  
   ~

   ~
   
   \begin{boldmath}
\subsubsection{Region II}
\end{boldmath}
   
      
      One can see from \eq{GSS101} and \eq{GSG3} that  $\Omega^{(0,II}$ is the solution of the equation:
     \beq \label{GSG5}
   \mathscr{U}\Lb\Omega^{(0,II)},  \Omega_0\Rb   \,\,=\,\, \sqrt{\frac{2}{\kappa}}\Lb \Lb 1 + \nu\Rb\,z \,+\,\nu \,t \,-\, \hat{z}\Rb
   \eeq
   where  $ \hat{z} =  \tilde{z} + 2 \nu \xi_{0,s}$ and  $\mathscr{U}$ is the same function that we have discussed in the previous section but with different values of the parameter $a$ (see \eq{GSG4}). The value of $\Omega_0$ (see \eq{GSG4} is large and therefore we can use \eq{GSS13}  which takes the form:
   
   \bea \label{GSG6}
&&   \Omega^{(0,II)}\Lb z,\xi_s \Rb\,-\,a\,\,=\\
&&\,\,\frac{ \Lb (1+\nu) z + \nu t - \hat{z}\Rb^2}{2\,\kappa} -  \sum^\infty_{k=1} \frac{(-1)^{k-1}(2k -1)!!}{2^k k\,k!}\Lb \frac{1}{\frac{\Lb (1+\nu) z + \nu t - \hat{z}\Rb^2 }{2\,\kappa}}\Rb^{k +\h} \exp\Lb - k\,\frac{ \Lb  (1+\nu) z + \nu t - \hat{z}\Rb^2}{2\,\kappa}\Rb \nn
   \eea
   or for    $ \Delta^{(0,II)}\Lb z,\xi_s \Rb $ we have  
   \beq \label{GSG7}  
    \Delta^{(0,II)}_0\Lb z ,\xi_s \Rb\,\,=\,\,\tilde{\Delta}_{LT}\Lb z,\xi_s\Rb\exp\Lb -\,\Omega_0  +  \sum^\infty_{k=1} \frac{(-1)^{k-1}(2k -1)!!}{2^k k\,k!}\Lb \frac{2 \kappa}{ \Lb (1+\nu) z + \nu t - \hat{z}\Rb^2}\Rb^{k +\h} \tilde{\Delta}^k_{LT}\Lb z, \xi_s\Rb  \Rb
    \eeq 
 where
   \beq \label{GSG8} 
    \tilde{\Delta}_{LT}\Lb z, \xi_s\Rb  \,\,=\,\,\exp\Lb - \frac{ \Lb (1+\nu) z + \nu t - \hat{z}\Rb^2}{2\,\kappa}  \Rb
    \eeq    
    It is instructive to find the solution restricting ourselves by the first term in the expansion of \eq{GSG7}:
    \beq \label{GSG9}
    \Delta^{(0,II)}_0\Lb z , \xi_s \Rb \,\,=\,\,\exp\Lb  - \frac{ \Lb (1+\nu) z + \nu t - \tilde{z}\Rb^2}{2\,\kappa}  -   
\frac{\Lb z_0 - \tilde{z}\Rb^2}{2\,\kappa} \,+\,\frac{\Lb z_0 -\xi_{0,s} - \tilde{z}\Rb^2}{\,\kappa}  \,-\,\h \exp\Lb  z_0 - \xi_{0,s} \Rb \Rb
\eeq

Recall, that $\nu $ has to be found from the boundary condition at $\xi = \xi^A_0$.  
 
 ~

 ~

 ~

 
  \begin{boldmath}
\subsubsection{Matching on the line $\xi=\xi^A_0$}
\end{boldmath}


The value of parameter $\nu$ in \eq{GSG7} (see also \eq{GSG9}),   can be found from  the matching conditions on line $\xi=\xi^A_0 $   or $z = z_A=\xi^A_0+ \kappa \dY $ and  $t=t_A = \kappa \dY - \xi^A_0$ has the following form:
  \bea \label{SDMAT1}
&&\Delta^{(0,II)}_0\Lb z_A , \xi_s \Rb \,\,= \\
&&\underbrace{\exp\Lb  - \frac{ \Lb (1+\nu) z _A+ \nu t _A-\, \tilde{z}- 2 \nu \,\xi_{0,s}\Rb^2}{2\,\kappa}  -   
\frac{\Lb z_{0,A} - \tilde{z}\Rb^2}{2\,\kappa} \,+\,\frac{\Lb z_{0,A} -\xi_{0,s} - \tilde{z}\Rb^2}{\,\kappa}  \,-\,\h \exp\Lb  z_{0,A} - \xi_{0,s} \Rb \Rb
}_{region~~II} \,\,\,\,\nn\\
&&=\,\,\underbrace{\Delta^{(0,I)}(\eta(z_A)) = C^2\exp\Lb - \frac{ \Lb z_A  - \tilde{z}\Rb^2}{2\,\kappa}\,\,-\,\, \Omega^{0,I}_0\Rb}_{region~~I}\nn\eea
  where $z_{0,A} = \xi^A_0+ \kappa \dY $.

  In the region I we use  the approximate solution to \eq{GSS17} (see \fig{om}). $\Omega^{0,I}_0  $ is determined by \eq{GSS50}. From \eq{SDMAT1}  one can  see that $\nu =0$ and 
  \beq \label{SDMAT2}
 C^2 = \exp\Lb \frac{\Lb \xi^A_0 - \tilde{z}\Rb^2}{\kappa} - \h e^{\xi^A_0}\Rb
  \eeq

     ~
  
  ~

\begin{boldmath}
\section{ $n^D$ in modified homotopy approach } 
\end{boldmath}

From \eq{DIFF} and \eq{AMPLITUD} we can calculate $n^D$ which is equal to
\beq \label{ND01}
 n^D\Lb z,Y_M, z_0,\dy; b \Rb
 \,\,=\,\, \frac{\pp N^D(z,\dY, \delta Y_0)}{\pp \dY}\,\,=\,\, - \frac{\pp\mathscr{ N}(z,\dY, \delta Y_0)}{\pp \dY}\,\,+\,\,2  \frac{\pp N(z,\dY, \delta Y_0)}{\pp \dY}
 \eeq
 However, before making this evaluation we return to the master equation of \eq{SDEQ} and rewrite it in the saturation region assuming  $N^D = 1 - \delta^D $, $  \mathscr{ N} = 1 - \Delta^D $ and $N = 1 - \Delta$. It takes the form
 \beq \label{ND02}
  \frac{\pp \delta^D(z,\dY, \delta Y_0)}{\pp \dY}\,\,=\,\, -\, \int \frac{d^2\,x_{02}}{2 \pi} \frac{ x^2_{01}}{x^2_{02}\,x^2_{12}}\Big\{  \Delta^D_{02} \Delta^D_{12} - 2 \Delta_{02}\Delta_{12}  \,+\, \delta^D_{01}\Big\}
  \eeq
  The linear approximation to \eq{ND02}  is
   \beq \label{ND03}  
 \frac{\pp \delta^D(z,\dY, \delta Y_0)}{\pp \dY}\,\,=\,\, -\int \frac{d^2\,x_{02}}{2 \pi} \frac{ x^2_{01}}{x^2_{02}\,x^2_{12}}\delta^D_{01}\,=\,-\,\z\,\delta^D_{01} 
 \eeq
 One can see that this is the same equation as for $\Delta^D$ or for  $\Delta$.  Hence in the linear approximation  we have in the region I:
    \bea \label{ND04}  
 n^D \,&=&\,z \exp\Lb -\frac{z^2 - z_0^2}{2 \kappa}\Rb\Bigg( 2 C\exp\Lb - \frac{z_0^2}{2 \kappa}\Rb  \,-\,C^2 \exp\Lb - \frac{z_0^2}{\kappa}\Rb\Bigg)\nn\\
 &=&\,z\, \exp\Lb -\Lb z + z_0\Rb Y_M\Rb\Bigg( 2 C\exp\Lb - \frac{z_0^2}{2 \kappa}\Rb  \,-\,C^2 \exp\Lb - \frac{z_0^2}{\kappa}\Rb\Bigg) 
\eea 
  and in the region II:
   \bea \label{ND05}   
   n^D \,&=&\,\z\, \exp\Lb -\frac{z^2 - z_0^2}{2 \kappa}\Rb \Bigg\{    
  2 G\Lb z_0 - \dy\Rb \exp\Lb - \frac{\Lb z_0 - \tilde{z}\Rb^2}{2\kappa}\Rb-   
  G^2\Lb z_0 - \dY\Rb \exp\Lb - \frac{\Lb z_0 - \tilde{z}\Rb^2}{\kappa}\Rb \Bigg\}
  \eea  
  
  In general case we can use \eq{ND01} and calculate $ \frac{\pp\mathscr{ N}(z,\dY, \delta Y_0)}{\pp \dY}$ using \eq{GSS101} and \eq{GSS17}. Indeed,
   \beq \label{ND06}
     \frac{ d \mathscr{U}\Lb\Omega^{(0,I)}\Rb}{d z} =  \frac{ d \mathscr{U}\Lb\Omega^{(0,I)}\Rb}{d \Omega^{(0,I)}}   \frac{ d \Omega^{(0,I)}}{d z} = 
     \frac{ d \Omega^{(0,I)}}{d z} \frac{1}{ \sqrt{ \Omega^{(0,I)} + \exp\Lb - \Omega^{(0,I)}\Rb - a }}    = \sqrt{\frac{2}{\kappa}}
     \eeq
     Therefore,
       \beq \label{ND07}     \frac{\pp\mathscr{ N}(z,\dY, \delta Y_0)}{\pp \dY} =
       \kappa \frac{ d \Omega^{(0,I)}}{d z} \exp\Lb - \Omega^{(0,I)}\Rb =   \sqrt{2\,\kappa} \, \sqrt{ \Omega^{(0,I)} + \exp\Lb - \Omega^{(0,I)}\Rb -a }\,\, \exp\Lb - \Omega^{(0,I)}\Rb
        \eeq  
        
        In the region I $a = \Lb z_0  - \tilde{z}\Rb^2/(2\kappa) - 2 \ln C$, while in the region II
        $a =  \,\,\frac{\Lb z_0 - \tilde{z}\Rb^2}{2\,\kappa} \,-\,\frac{\Lb z_0 -\xi_{0,s} - \tilde{z}\Rb^2}{\,\kappa}  \,+\,\h \exp\Lb  z_0 - \xi_{0,s} \Rb$ and we need to replace $\Omega^{(0,I)} $ by $\Omega^{(0,II)} $.
  
  Actually, the same formula is correct for the elastic amplitude since \eq{SDBK} coincides with the BK equation \cite{BK}. The  only difference in the values of $a$. In region I $a$ for the elastic amplitude is equal to
  $a =\Lb z_0 - \tilde{z}\Rb^2/(2 \kappa) -  \ln C$    and in the region II $a =   \,\,\frac{\Lb z_0 - \tilde{z}\Rb^2}{2\,\kappa} \,-\,\frac{\Lb z_0 -\xi_{0,s} - \tilde{z}\Rb^2}{2\,\kappa}  \,+\,\frac{1}{4} \exp\Lb  z_0 - \xi_{0,s} \Rb$. 
  
  ~

  ~

 ~

\begin{boldmath}
\section{The second  iteration for $\Delta^D$ in the modified homotopy approach} 
\end{boldmath}
  In this section we return to the discussion of \eq{SDBK12} for which we used a particular contribution. Now we are going to insert in this term $\Delta^D_{0,1}\,\,=\,\,\Delta^D_0$ and develop the second iteration of the master equation, choosing $ \mathscr{N_{L}}[\Delta^D]  $ in the form
 \beq \label{2I1}
\mathscr{N_{L}}[\Delta^D]\,\,=\,\,\bas\int \frac{d^2\,x_{02}}{2 \pi} \frac{ x^2_{01}}{x^2_{02}\,x^2_{12}} \Delta^D_{0} \Lb x_{02}\Rb \Delta^D_{0}\Lb x_{12}\Rb -  \Delta^D_{0} \intl^{x^2_{01}} \frac{d x^2_{02}}{x^2_{02}}  \Delta^D_{02}\eeq

Let us estimate the first term in this equation  plugging in  solution $\Delta^D_0$ at large $z$, viz.:

  \beq \label{2I2}     
         \Delta^D_0\Lb z, \xi\Rb\,\,=\,\,\exp\Lb  - \frac{\Lb z -   \tilde{z} \Rb^2}{2 \kappa}\,+\,\phi^I\Lb \zeta, z_0\Rb \Rb\,\,\xrightarrow\,\,\exp\Lb  - \frac{ z^2}{2 \kappa}\,+\,\phi^I\Lb \zeta, z_0\Rb \Rb      \eeq

\begin{figure}
 	\begin{center}
 	\leavevmode
 		\includegraphics[width=12cm]{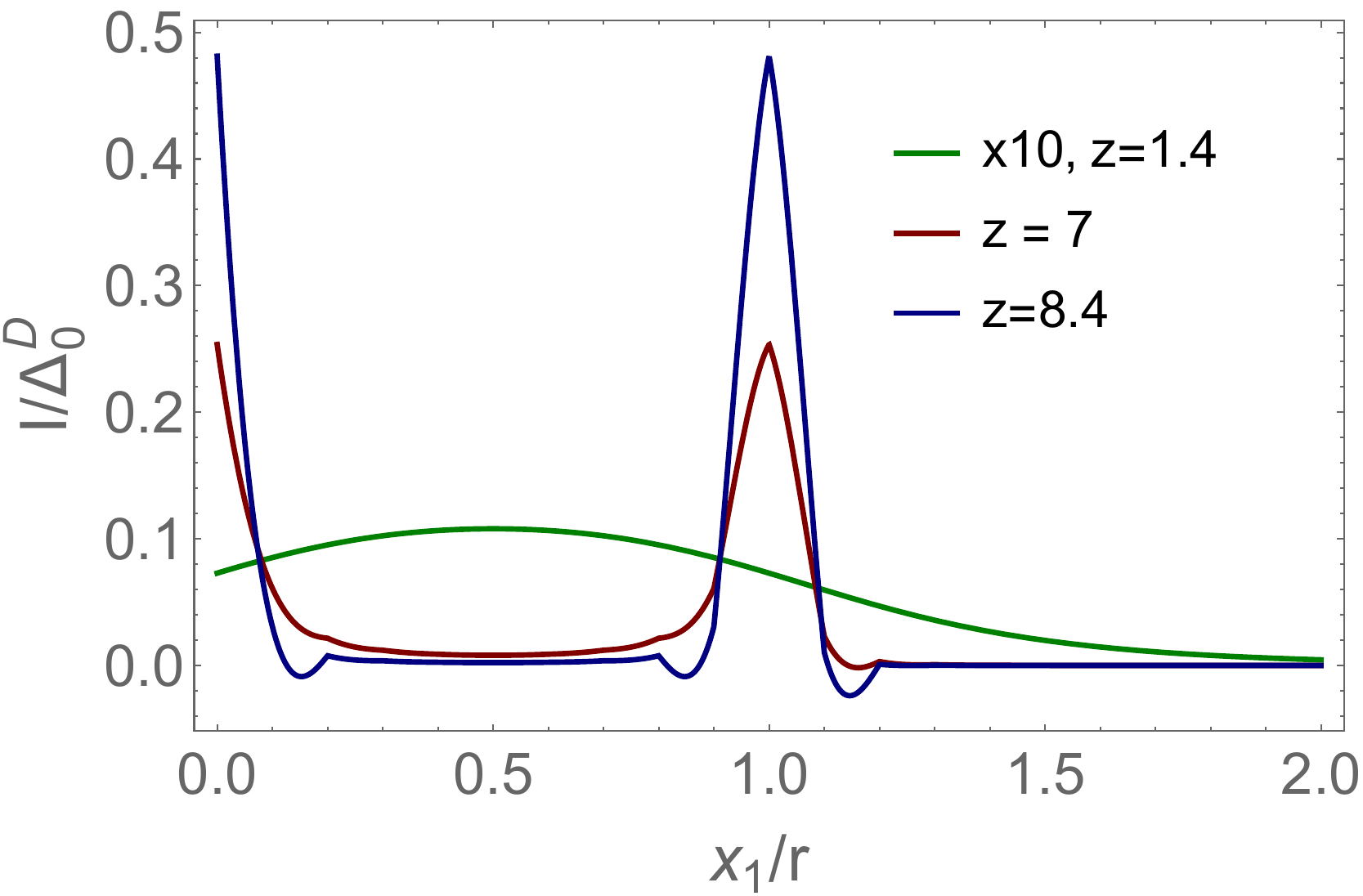}
 	\end{center}
 	\caption{ The ratio  $I\Lb \vec{r}, x_1\Rb/\Lb \Delta^D_0\Lb z\Rb\Rb$  from \eq{2I3}   versus  $x_1/r$ at different values of $z$.}
 	\label{ivsx}
 \end{figure}

Denoting $\vec{x}_{01} $ by $\vec{r}$ and introducing $\vec{x}_{02} = \Lb x_1,x_2\Rb$ where $x_1$ is the projection on the direction of $\vec{r}$, we can rewrite the integral in the following form:
   \bea \label{2I3}
&&\bas\int \frac{d^2\,x_{02}}{2 \pi} \frac{ x^2_{01}}{x^2_{02}\,x^2_{12}} \Delta^D_{02} \Delta^D_{12} \,\,=\,\, \int d x_{1} \,I\Lb \vec{r}, x_1\Rb\\
&&=e^{ 2 \phi^I\Lb \zeta, z_0\Rb} \int \frac{r^2\,\,d x_1 d x_2}{\Lb x^2_1 + x^2_2\Rb \, \Lb \Lb r - x_1\Rb^2 + x^2_2\Rb } \exp\Lb - \frac{1}{2 \kappa} \Bigg( \ln^2\Lb Q^2_s\Lb x^2_1 + x^2_2\Rb\Rb  
+ \ln^2\Lb Q^2_s\Lb ( r - x_1)^2 + x^2_2\Rb\Rb \Bigg)\Rb\nn\\
&& =e^{ 2 \phi^I\Lb \zeta, z_0\Rb} \int \frac{r^2\,\,d x_1 d x_2}{\Lb x^2_1 + x^2_2\Rb \, \Lb \Lb r - x_1\Rb^2 + x^2_2\Rb } \exp\Lb -  \Psi\Rb\nn
\eea
where $Q^2_s\Lb x^2_1 + x^2_2\Rb\,\,>\,\,1$ and $ Q^2_s\Lb ( r - x_1)^2 + x^2_2\Rb\,\,>\,\,1$. In  the region where these arguments are large we can use the method of steepest descent to take the integral. The equation for the saddle point are
\begin{subequations} 
\bea
\frac{\pp \Psi}{\pp\,x_2}\,=\,0&\to & \frac{1}{\kappa} \Bigg\{ \ln\Lb Q^2_s\Lb x^2_1 + x^2_2\Rb\Rb\frac{2 x_2}{x^2_1 + x^2_2}\,\,+\,\,\ln\Lb Q^2_s\Lb (r - x_1)^2 + x^2_2\Rb\Rb\frac{2 x_2}{(r - x_1)^2 + x^2_2}\Bigg\}\,\,=\,\,0;\label{2I31}\\
\frac{\pp \Psi}{\pp\,x_1}\,=\,0&\to&  \frac{1}{\kappa} \Bigg\{ \ln\Lb Q^2_s\Lb x^2_1 + x^2_2\Rb\Rb\frac{2 x_1}{x^2_1 + x^2_2}\,\,-\,\,\ln\Lb Q^2_s\Lb (r - x_1)^2 + x^2_2\Rb\Rb\frac{2 (r - x_1)}{(r - x_1)^2 + x^2_2}\Bigg\}\,\,=\,\,0;\label{2I32}
\eea
\end{subequations} 
One can see that \eq{2I31} does not have solution. It means that main contribution stems from $x_2  \to 0$ which is the end point of the integral.
\eq{2I32} has the solution $x_1 = \h r$. So we can see two main contributions in the integral over $x_1$: the end  point of the integral: $x_1 \to 0$ and the contribution of the saddle point. The first contribution has been taken into account in \eq{SDBK12}  and it has been subtracted in \eq{2I1}. 
The estimates using the method of steepest descent has to be discussed calculating the second derivatives in the saddle point $\vec{x}_{SP} = \Lb \h r, 0\Rb$. 
\begin{subequations} 
\bea
\frac{\pp^2 \Psi}{\pp\,x_2^2}\Bigg{|}_{\vec{x}_{02} =\vec{x}_{SP}}\,&= & \,-\,\frac{16}{\kappa}  \ln\Lb Q^2_s\, r^2/4\Rb;\label{2I04}\\
\frac{\pp^2 \Psi}{\pp\,x_1^2}\Bigg{|}_{\vec{x}_{02} =\vec{x}_{SP}}&=& \,-\,\frac{16}{\kappa}\Lb 2  -    \ln\Lb Q^2_s\, r^2/4\Rb\Rb;\label{2I05}
\eea
\end{subequations} 
From 
\eq{2I05} one can see that the saddle point approximation can be justified only in the limited region of $z$: $ 2 + 2 \ln 2 \, \equiv z_{max} \,> \,z\, >0$. For large $z > 0$  the second derivatives is positive indicating that the integral has a  minimum in this point. Integrating over $x_2$ we plot in \fig{ivsx} $I\Lb \vec{r}, \vec{x}_{02}\Rb/\Lb \Delta^D_0\Lb z\Rb\Rb^2$ versus $x_1/r$ and different values of $z$. One can see that at $z < z_{max}$ the integrant has a maximum at $x_1/r = 1/2$ and can be taken using the method of steepest descent. However for $z > z_{max}$ this maximum disappears and only regions $x_{02} \to 0$ and $x_{02} \to r$ contribute. These contributions have been taken into  account in our first iteration. Hence we expect that the corrections from the second iteration can be small.

\begin{figure}
 	\begin{center}
 	\leavevmode
 		\includegraphics[width=12cm]{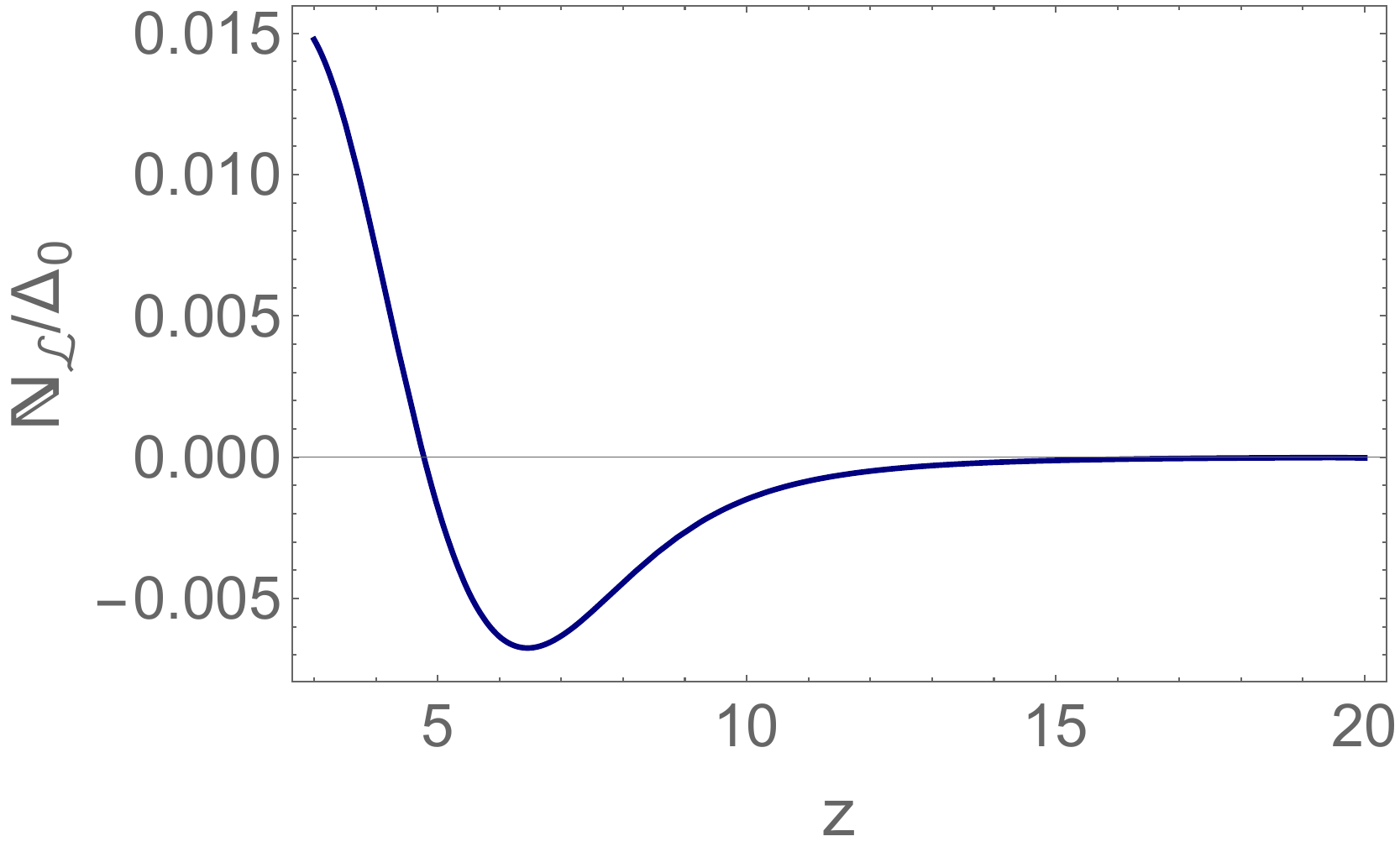}
 	\end{center}
 	\caption{The ratio $ \mathscr{N_{L}}[\Delta^D_{0}]/\Delta^D_0$   versus $z$ for $z > z_0$. $z_0$ is taken to be 3. $\bas = 0.2$}
 	\label{d2}
 \end{figure}

  In \fig{d2} we present the numerical estimates for       
$\mathscr{N_{L}}$ of \eq{HOM1} :
 \beq \label{2I5}
 \mathscr{N_{L}}[\Delta^D_{0}]/\Delta^D_0\,\,=\,\,\Lb \bas \int \frac{d^2\,x_{02}}{2 \pi} \frac{ x^2_{01}}{x^2_{02}\,x^2_{12}} \Delta^D_{0} \Lb x_{02}\Rb \Delta^D_{0}\Lb x_{12}\Rb\,\,-\,\,\Delta^D_{0} \Lb z \Rb \int^z_0 d z'\Delta^D_{0} \Lb z' \Rb\Rb/\Delta^D_0 \eeq
One can see that for $z>z_0$ $ \mathscr{N_{L}}$ turns out to be small.

 ~
 
 ~

\begin{boldmath}
\subsection{Geometric scaling solution} 
\end{boldmath}
 
 The equation for the second iteration $\Delta^D_1
 $ takes the form (see \eq{HOM2}):
  \beq \label{2I60} 
  \Bigg( \kappa \frac{\pp\,}{\pp\,z}\,\,+\,\,z\,\,-\,\, \zeta\Bigg) \Delta^D_1\Lb z, z_0\Rb\,+\,\,\underbrace{\Delta^D_0\Lb z,z_0\Rb \intl^\infty_z  d z'  \Delta^D_{1} \Lb z',\z_0\Rb\,+\,\,\Delta^D_1\Lb z,z_0\Rb \intl^\infty_z  d z'  \Delta^D_{0} \Lb z',\z_0\Rb}_{\sim \Lb\Delta^0\Rb^3}=\,\,-  \underbrace{ \mathscr{N_{L}}[\Delta^D_{0}\Lb z\Rb]}_{\sim \Lb\Delta^0\Rb^2} 
 \eeq
 Taking into account only terms of the order of $\Lb\Delta^0\Rb^2$, we have
  \beq \label{2I6} 
\Bigg( \kappa \frac{\pp\,}{\pp\,z}\,\,+\,\,z\,\,-\,\, \zeta\Bigg) \Delta^D_1\Lb z, z_0\Rb\,\,=\,\,- \mathscr{N_{L}}[\Delta^D_{0}\Lb z\Rb]
 \eeq

\begin{figure}
 	\begin{center}
 	\leavevmode
 		\includegraphics[width=12cm]{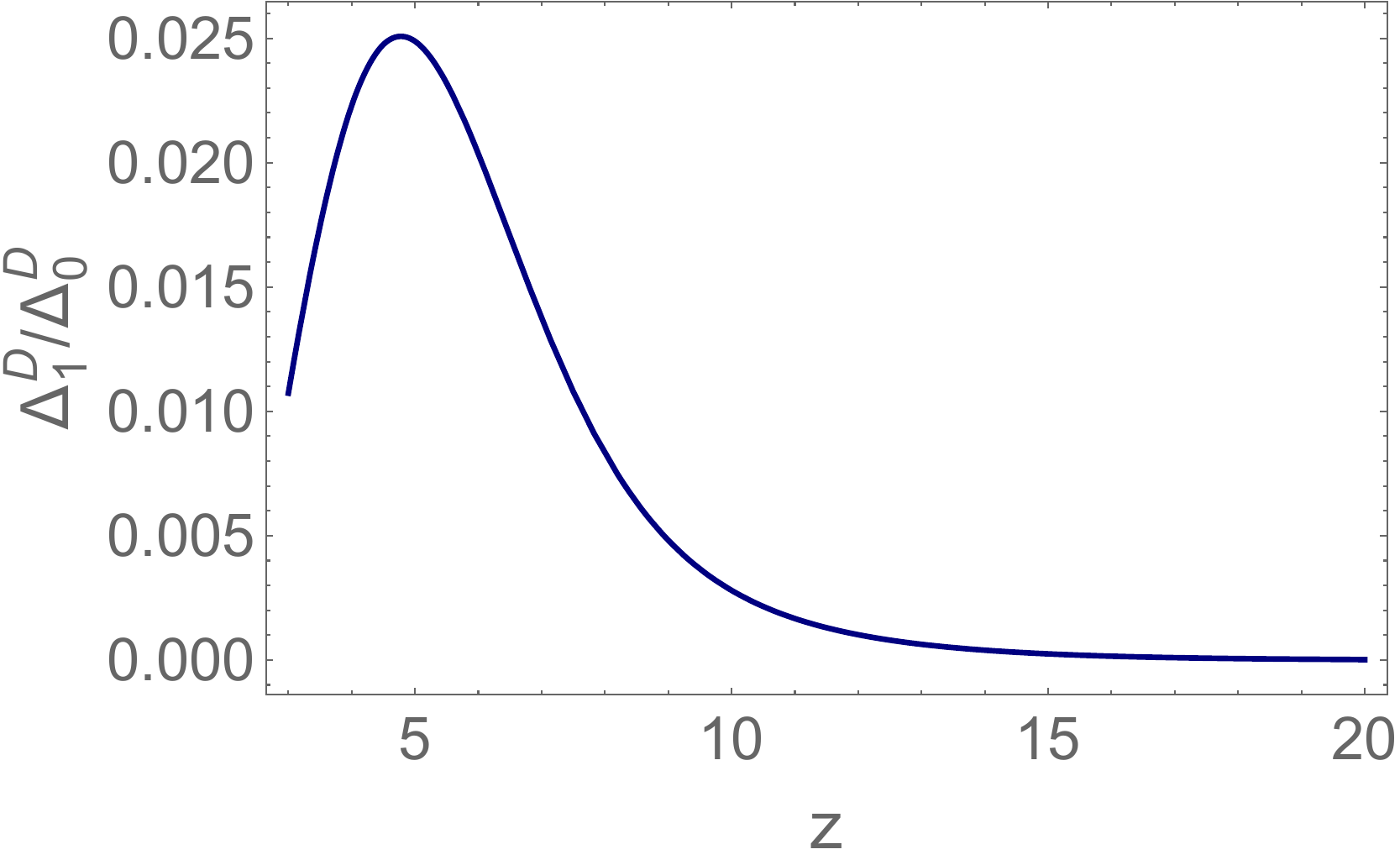}
 	\end{center}
 	\caption{The ratio $\Delta^D_{1}/\Delta^D_0$   versus $z$ for $z > z_0$. $z_0$ is taken to be 3. $\bas = 0.2$}
 	\label{d0d1}
 \end{figure}

 The particular solution can be written as follows
 \beq \label{2I7} 
 \Delta^D_1\Lb z, z_0\Rb\,\,=\,\,- \Delta^D_0\Lb z, z_0\Rb \intl^\infty_z\,d z' \frac{1}{ \Delta^D_0\Lb z', z_0\Rb}\,\mathscr{N_{L}}[\Delta^D_{0}\Lb z'\Rb]
 \eeq
 As we have discussed (see \eq{2I5}) we expect that  $ \Delta^D_1\Lb z, z_0\Rb$ will be small. Indeed, \fig{d0d1} shows that the ration $\Delta^D_1\Lb z, z_0\Rb/\Delta^D_0\Lb z, z_0\Rb$ turns out to be small. 
           
    We need to add the general solution to the linear equation to satisfy the initial condition , which takes the form:
    \beq \label{ICD1}
 \Delta^D_1\Lb z = z_0, z_0\Rb\,\,=\,0 \eeq
   Hence the general form for $ \Delta^D_1\Lb z, z_0\Rb$ is
    \beq \label{3ID310} 
    \Delta^D_1\Lb z, z_0\Rb\,= \,\,  \underbrace{\Delta^D_1\Lb z, z_0\Rb  }_{\rm\eq{2I7}} \,\,+\,\,\underbrace{C_2 \Delta^D_0\Lb z, z_0\Rb}_{\rm solution\,\,to\,\,linear\,\,equation}
      \eeq

    Finding constant ${\rm C_2}$ from \eq{ICD1} one can see that this constant turns out to be rather small, since $\Delta^D_1\Lb z, z_0\Rb$ turns put to be small at $z =z_0$ (In  \fig{d0d1}   ${\rm C_2} = 0.017$ at $z_0 =3$).

       ~
       
       ~

\begin{boldmath}
\subsection{Region II} 
\end{boldmath}

In  region  II  (see \fig{sat}) the elastic scattering amplitude does not have the geometric scaling behaviour being the function of two variables: $z_0$ and $\xi$. This function has been found in our previous paper \cite{CLMNEW} 
(see Eq.58 and Eq.67) and it has the form of \eq{RIIEL1} for the initial conditions for $\Delta^{0,II}_0$.  Bearing this in mind, 
   we choose $\mathscr{L}$ in homotopy approach in the form of \eq{LBK} in  region II. The chosen $\mathscr{L}$   gives 
    \eq{2I1} for $\Delta^{(0,II)}_0$.  For region II  $\mathscr{N_{L}}[\Delta^D] $ has the same form as of \eq{2I1}
  \beq \label{2II1}
\mathscr{N_{L}}[\Delta^{(0,II)}_0]\,\,=\,\,\bas\int \frac{d^2\,x_{02}}{2 \pi} \frac{ x^2_{01}}{x^2_{02}\,x^2_{12}} \Delta^{(0,II)}_0 \Lb x_{02}\Rb \Delta^{(0,II)}_0\Lb x_{12}\Rb -  \Delta^{(0,II)}_0 \intl^{x^2_{01}} \frac{d x^2_{02}}{x^2_{02}} \Delta^{(0,II)}_0\eeq

     We need to investigate $\mathscr{N_{L}}[\Delta^{(0,II)}_0]$ in the same way as we did in \eq{2I1}-\eq{2I3}. For this analysis we take a simplified version of $ \Delta^{(0,II)}_0 $ of \eq{SDMAT2} and \eq{2I2}. This version  stems from the solution of the linear equation in which we neglect the non-linear terms in \eq{SDBK2}.     
         
     The solution to this linear equation is
       \beq \label{RIID02}     
        \Delta^D_0\Lb z, \xi,z_0\Rb\,\,=\,\,\exp\Lb - \frac{z^2}{2 \kappa} \,\,+\,\,\phi^{II}\Lb \xi,z_0\Rb\Rb
        \eeq
        Function $\phi^{II}\Lb \xi,z_0\Rb$ has to be found from the initial condition of \eq{RIIEL1}, viz.:
        \beq \label{RIID03}
\exp\Lb \phi^{II}\Lb\xi,  z_0\Rb\Rb\,\,=\,\,\exp\Lb  \frac{z_0^2}{2 \kappa}\Rb
G^2\Lb \xi\Rb \exp\Lb - \frac{\Lb z_0-\tilde{z}\Rb^2}{\kappa} \Rb
       \eeq

From \eq{RIID03} 
\beq \label{RIID04}
\phi^{II}\Lb\xi,  z_0\Rb\,\,=\,\,\frac{z^2_0}{2 \kappa} - \frac{\Lb z_0-\tilde{z}\Rb^2}{\kappa} \,+\,\frac{\Lb \xi -\tilde{z}\Rb^2}{\kappa} - \h e^\xi.
\eeq
   Using \eq{RIID04} we are going to discuss the first term in \eq{2II1}.
  From \eq{RIID02} and \eq{RIID04} we can conclude that the main contributions at large dipole sizes stem from the factor $\h e^\xi$. This factor changes the   behaviour of the integrant in \eq{2II1} and the integral can be taken in the method of steepest descent at $\vec{x}_{02} = \h \vec{x}_{01} \equiv  \h \vec{r}$. In particular the contribution of the small size dipoles, that was essential in the region I , turns out to be small.

   Let us demonstrate all of these features. First we rewrite the solution of \eq{RIID02} as follows:
      \beq \label{RIID12}   
      \Delta^D_0\Lb z,\xi, z_0\Rb=\exp\Lb - \Sigma\Lb z(\xi),\xi,z_0(\xi)\Rb \,\,-\,\,\h e^\xi\Rb~~\mbox{with}~~\Sigma\Lb z,\xi,z_0\Rb\,\,=\,\,\frac{z^2}{2 \kappa} - \frac{z^2_0}{2  \kappa} +\frac{( z_0 -\tilde{z})^2}{\kappa}  -  \frac{( \xi -\tilde{z})^2}{\kappa} ,      \eeq
plugging this equation in $\mathscr{N_{L}}[\Delta^D] $ we have
    \bea \label{RIID13}
   &&\mathscr{N_L}\,\,=\,\,\bas\int \frac{d^2\,x_{02}}{2 \pi} \frac{ x^2_{01}}{x^2_{02}\,x^2_{12}} \Delta^D_0\Lb x_ {02}\Rb \Delta^D_0\Lb x_{12}\Rb \,\, =\\
   &&= \int  \frac{d^2\,x_{02}}{2 \pi} \frac{ x^2_{01}}{x^2_{02}\,x^2_{12}} \exp\Lb - \Sigma\Lb z(\xi_{02}),\xi_{02},z_0(\xi_{02})\Rb  - \Sigma\Lb z(\xi_{12}),\xi_{12},z_0(\xi_{12})\Rb\,\,-\,\,\h Q^2_s \Lb x^2_{02}+ x^2_{12}\Rb\Rb\nn,  \eea    
 where we denote $\xi_{ik} = \ln\Lb Q^2_s\Lb \dY=0\Rb) x^2_{ik} \Rb$, $z\Lb \xi_{ik}\Rb = \kappa \dY + \xi_{ik}$ and $ z_0\Lb \xi_{ik}\Rb = \kappa  \delta Y_0 + \xi_{ik}$. Introducing $\vec{x}_{02} = \h \vec{r} + \vec{x}$ and $ 
  \vec{x}_{12} = \h \vec{r} - \vec{x}$ we obtain:
   \bea \label{RIID14}
   &&\mathscr{N_L}\,\,=\,\,\bas\intl  \frac{d^2\,x_{02}}{2 \pi} \frac{ x^2_{01}}{x^2_{02}\,x^2_{12}}  \Delta^D_0\Lb x_ {02}\Rb \Delta^D_0\Lb x_{12}\Rb \,\, =\\
   &&= \bas\intl  \frac{d^2\,x}{2 \pi} \frac{ r^2}{(\frac{1}{4} r^2 +  x^2)^2 -\Lb \vec{r} \cdot \vec{x}\Rb^2} \exp\Lb - \Sigma\Lb z(\xi_{02}),\xi_{02},z_0(\xi_{02})\Rb  - \Sigma\Lb z(\xi_{12}),\xi_{12},z_0(\xi_{12})\Rb\,\,-\,\,\h Q^2_s \Lb \h r^2 + 2\,x^2 \Rb\Rb\nn  \eea   
  One can see that the main contribution stems from $d x \sim1/Q_s$ since  functions $\Sigma$ are logarithmic. Taking this integral for such $x$ we derive
   \bea \label{RIID15}
  \mathscr{N_L}\,\,&=&\,\, 2\, \bas e^{-\xi_r} \exp\Lb- 2 \Sigma\Lb z(\xi_r),\xi_r,z_0(\xi_r)\Rb - e^{\xi_r}\Rb\,\nn\\
  &=&\,2\, \bas e^{-\xi_r} \exp\Lb -2\Bigg(\frac{z(\xi_r)^2}{2 \kappa} - \frac{z_0(\xi_r)^2}{2  \kappa} +\frac{( z_0(\xi_r) -\tilde{z})^2}{\kappa}  -  \frac{( \xi_r -\tilde{z})^2}{\kappa}\Bigg)   - e^{\xi_r}\Rb\nn\\
  &=& \bas \exp\Lb - \frac{\Lb z - 2 \ln2\Rb^2}{\kappa} \,+\,\tilde{\phi}\Lb \xi_r,z_0\Rb\Rb 
\eea
  with $\tilde{\phi}\Lb \xi_r,z_0\Rb=2\phi^{II}\Lb \xi_r,z_0\Rb-\xi_r+\ln 2$,  $\xi_r = \ln \Lb Q^2_s\,r^2/4\Rb  =  \xi - 2 \ln 2$, $z(\xi_r)  = z - 2 \ln 2$ and $z_0(\xi_r)  = z_0 - 2\ln 2$. 
  
  The equation for the second iteration  takes the form (see \eq{HOM2}):
  \beq \label{RIID16} 
 \Bigg( \kappa \frac{\pp\,}{\pp\,z}\,\,+\,\,z\Bigg) \Delta^D_1\Lb z, \xi, z_0\Rb\,\,=\,\,-   \mathscr{N_{L}}[\Delta^D_{0}\Lb z\Rb]\,\,=\,\,-\exp\Lb - \frac{\Lb z - 2 \ln2\Rb^2}{\kappa} \,-\,\tilde{\phi}\Lb \xi,z_0\Rb\Rb  \eeq  
 Note, the two  changes we make in \eq{SDBK2}: first, we use the variables $z$ and $\xi$ instead of $z$ and $\dY$; and second, we neglect the non-linear contributions by the same reason as in \eq{2I6}.

  The particular solution to this equation has the form:
   \bea \label{RIID17}  
   \kappa\, \Delta^D_1\Lb z, \xi, z_0\Rb&=&\Delta^D_0\Lb z, \xi, z_0\Rb\int^\infty_z d z' \frac{  \mathscr{N_{L}}[\Delta^D_{0}\Lb z'\Rb]}{\Delta^D_0\Lb z, \xi, z_0\Rb} \nn\\
    & =& \Delta^D_0\Lb z, \xi, z_0\Rb\int^\infty_z d z'\exp\Lb - \frac{\Lb z' - 2 \ln2\Rb^2}{\kappa} \,+\, \frac{\z'^2}{2\,\kappa}\,+\,(\tilde{\phi}\Lb \xi_r,z_0\Rb-{\phi}^{II}\Lb \xi_r,z_0\Rb) \Rb \nn\\
    &=&\Delta^D_0\Lb z, \xi, z_0\Rb\sqrt{\frac{\pi  \kappa}{2}}  e^{\frac{4\ln^2 2}{\kappa }} \text{erfc}\left(\frac{z- 4 \ln2}{\sqrt{2 \kappa}}\right) e^{ \,(\tilde{\phi}\Lb \xi_r,z_0\Rb-\phi^{II}\Lb \xi_r,z_0\Rb)}  \xrightarrow{z\,\gg\,1} \frac{\kappa}{z} \Lb \Delta^D_0\Lb z, \xi, z_0\Rb\Rb^2       \eea
    
    Finally, the solution which satisfies  \eq{ICD1} has the following form:
    \bea \label{RIID18}  
   \kappa\, \Delta^D_1\Lb z, \xi, z_0\Rb\,\,&=&\,\, \Delta^D_0\Lb z, \xi, z_0\Rb\int^z_{z_0} d z' \frac{  \mathscr{N_{L}}[\Delta^D_{0}\Lb z'\Rb]}{\Delta^D_0\Lb z, \xi, z_0\Rb} \nn\\
    &  \,=\,&\Delta^D_0\Lb z, \xi, z_0\Rb\sqrt{\frac{\pi  \kappa}{2}}  e^{\frac{4\ln^2 2}{\kappa }} \left(-\text{erf}\left(\frac{z_0 - 4\ln2}{\sqrt{2 \kappa }}\right)+\text{erf}\left(\frac{z- 4\ln2}{\sqrt{2 \kappa}  }\right)\right)         \eea
    
    \eq{RIID18} has the same qualitative features as $\Delta^D_1$ in region I: 
    it vanishes at $z \to z_0$ and in the region of large $z$ it approaches 
    ${\rm C}\,\, \Delta^D_0$ with small constant ${\rm C}\,\,\sim\,\,\Delta^D_0\Lb z_0,\xi,z_0\Rb/z_0$.
    
   ~
   ~

    ~
    
    ~
\begin{boldmath}
\section{Modified homotopy approach for $\Delta^D$ in perturbative QCD  region } 
\end{boldmath}

In this section we consider the perturbative QCD (pQCD)  region for the elastic amplitude. This region corresponds to the following kinematic restriction: 
$Q_s\Lb \dY - \dy\Rb\,r \geq 1$  but $ Q_s\Lb\dy\Rb\,r \leq1$.  $\dy$ is so small that we can use in this region the BFKL Pomeron exchange for the scattering amplitude. First we consider the diagram of \fig{gen1}-b, the expression for which has been written in \eq{DIFF21} with the only difference that $N_{el} =N^{BFKL}_{el}$ is described by the exchange of the BFKL Pomeron, viz.:
  \beq \label{DIFP1}
  n^D\Lb \fig{gen1}-b\Rb = \intl^{\epsilon + i \infty}_{\epsilon - i \infty}\frac{ d \gamma}{2 \pi i}\exp\Lb \bas \chi\Lb \gamma\Rb Y_M\Rb \phi_\gamma\Lb \vec{r}_{\perp} , \vec{r}^{\prime}, \vec{b}\Rb \Lb  N^{BFKL}\Lb \dy, r'\Rb\Rb^2 \frac{d^2 r^{\prime}  d^2 b }{4\,\pi^2\,r'^4}
  \eeq
Assuming that we are in the vicinity of the saturation scale for the elastic amplitude, we see that \eq{DIFP1} takes the form:
 \beq \label{DIFP2}
  n^D\Lb \fig{gen1}-b\Rb = \intl^{\epsilon + i \infty}_{\epsilon - i \infty}\frac{ d \gamma}{2 \pi i}\exp\Lb \bas \chi\Lb \gamma\Rb Y_M\Rb \phi_\gamma\Lb \vec{r}_{\perp} , \vec{r}^{\prime}, \vec{b}\Rb \Lb  r'^2\, Q^2_s\Lb \dy\Rb\Rb^{2 \bar{\gamma}} \frac{d^2 r^{\prime}  d^2 b }{4\,\pi^2\,r'^4}
  \eeq
  From  \eq{EIGENF} one can see that $ \int\frac{ d^2 b }{2 \,\pi} \phi_\gamma\Lb \vec{r}_{\perp} , \vec{r}^{\prime}, \vec{b}\Rb  =\frac{1}{2(2 \gamma - 1)}\,r'^2\, \Lb \frac{ r^2}{r'^2}\Rb^\gamma$.
Since $\gamma <1$,  one can see that the  highest values of $r'$ in $N_{el}$ contribute to the integral over $r'$.  This highest value is $r'^2 = 1/Q^2_s\Lb \dy\Rb$. Hence
 \beq \label{DIFP3}
  n^D\Lb \fig{gen1}-b\Rb = \intl^{\epsilon + i \infty}_{\epsilon - i \infty}\!\!\!\frac{1}{4 (2 \gamma -1) \gamma }\frac{ d \gamma}{2 \pi i}\exp\Lb \bas \chi\Lb \gamma\Rb Y_M\Rb \phi_\gamma\Lb \vec{r}_{\perp} , 1/Q_s\Lb \dy\Rb, \vec{b}\Rb \,\,\propto \,\,\Lb r^2\,Q^2_s\Lb \dY\Rb\Rb^{\bar{\gamma}}  \eeq 
Taking the integral over $\gamma$ using the method of steepest decent, we see that 
 \beq \label{DIFP3}
  n^D\Lb \fig{gen1}-b\Rb = n^D_0\Lb r^2\,Q^2_s\Lb \dY\Rb\Rb^{\bar{\gamma}} 
  \eeq
  
  In other words, the cross section of diffraction production is a  finite part of the total cross section\cite{MM1}\footnote{ In Ref.\cite{MM1} $\sigma_{diff}/\sigma_{tot} \propto \Lb \frac{\dY}{\dy\,(\dY - \dy)} \Rb^{3/2}$ . This relation comes from the previous estimates if we took into account that the saturation scale has the general form\cite{MUT}  $Q_s\Lb \dY\Rb = Q_s\Lb \eq{QS}\Rb  \Lb \frac{1}{\dY}\Rb^{\dfrac{3}{2 \bar{\gamma}} }$. }

\eq{DIFP3} allows us to find the boundary and initial conditions for $N^D$: for initial condition we have
 
\beq \label{DIFPIC}
\mathscr{N}^D(z \to z_0,\dY = \delta Y_0 )=2\,N^{\rm BFKL}\Lb z_0,\delta Y_0\Rb\,-\, \Lb N^{\rm BFKL}\Lb z_0,\delta Y_0\Rb \Rb^2 =2\,N_0 \Lb r^2\, Q^2_s\Lb \dy\Rb\Rb^{ \bar{\gamma}}-N^2_0 \Lb r^2\, Q^2_s\Lb \dy\Rb\Rb^{2 \bar{\gamma}}\eeq
which can be rewritten for $\Delta^D$ in the form:
\beq \label{DIFPIC1}
\Omega^{0,pQCD}_0\,\,=\,\,\mathscr{N}^D(z \to z_0,\dY = \delta Y_0  , \delta Y_0)\,\,=2\,N_0 \Lb r^2\, Q^2_s\Lb \dy\Rb\Rb^{ \bar{\gamma}}-N^2_0 \Lb r^2\, Q^2_s\Lb \dy\Rb\Rb^{2 \bar{\gamma}} \,\,=\,\,2 N_0\,e^{\bar{\gamma}\,z_0} - N^2_0e^{2\,\bar{\gamma}\,z_0}\eeq
with $z_0 \leq 0$.

 The boundary conditions from \eq{DIFP3} takes the form:
 \beq \label{DIFPBC} 
 \dfrac{\pp\ln\Lb N^D\Rb}{\pp\,z}\Bigg{|}_{z \to z_0,\dY = \delta Y_0}\,\, =
  \dfrac{\pp\ln\Lb \Omega^D\Rb}{\pp\,z}\Bigg{|}_{z \to z_0,\dY = \delta Y_0} 
 \,\,=\,\,\bar{\gamma}
 \eeq
 if we assume that $N_0$ is small.
 
 \begin{figure}
 	\begin{center}
 	\leavevmode
 		\includegraphics[width=7cm]{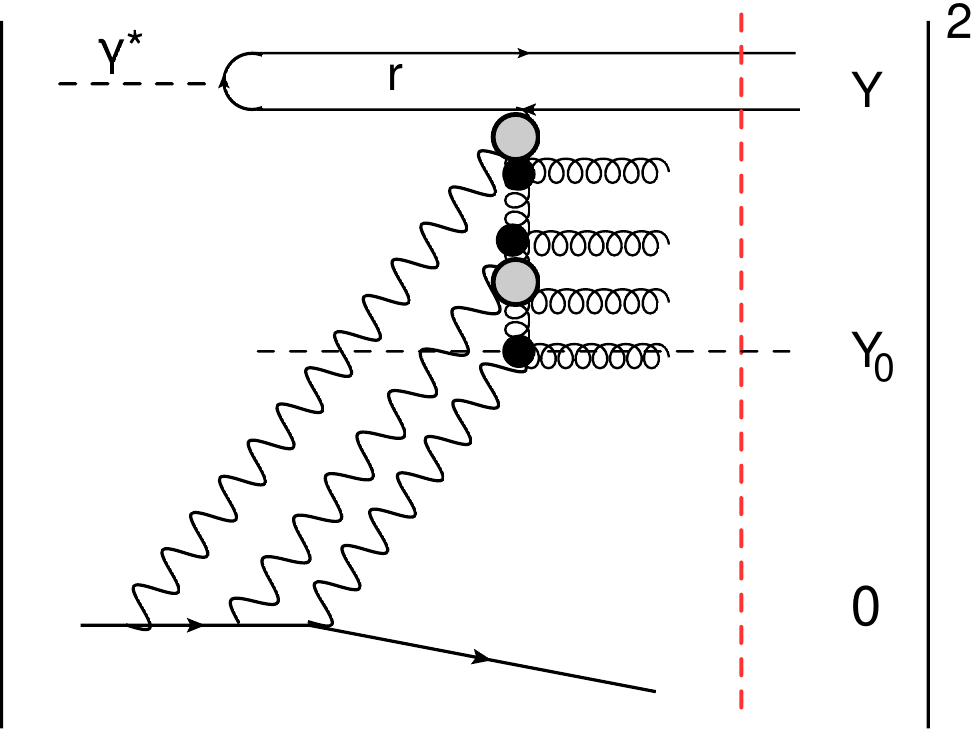}
 	\end{center}
 	\caption{  The diffractive dissociation in perturbative QCD region.}
 	\label{genp}
 \end{figure}        
       

 Therefore, we   need to solve \eq{GSS2} with the initial and boundary conditions given by \eq{DIFPIC1} and \eq{DIFPBC}. The general solution is known ( see \eq{GSS5} ) but we have to choose coefficients $C_1$ and $C_2$ to satisfy them. It should be noted that in perturbative QCD region for the elastic amplitude we are looking for the solution with the geometric scaling behavior (see \fig{sat}).
 The value of $C_2$ is $- z_0$ in this region, while $C_1$ can be found from the equation which follows from \eq{GSS5} by differentiation:
  \beq \label{DIFP4} 
  \dfrac{d\, \Omega^{0,pQCD} }{ d z}\Bigg{|}_{z=z_0}\Bigg{/} \Omega^{0,pQCD}_0= \sqrt{ \frac{1}{\kappa}\Lb \Omega^{0,pQCD}_0\Rb^2\,+\, C_1}\,\,=\,\,\bar{\gamma}
  \eeq  
 \begin{figure}
 	\begin{center}
	\begin{tabular}{ c c}
 	\leavevmode
 		\includegraphics[width=7.7cm]{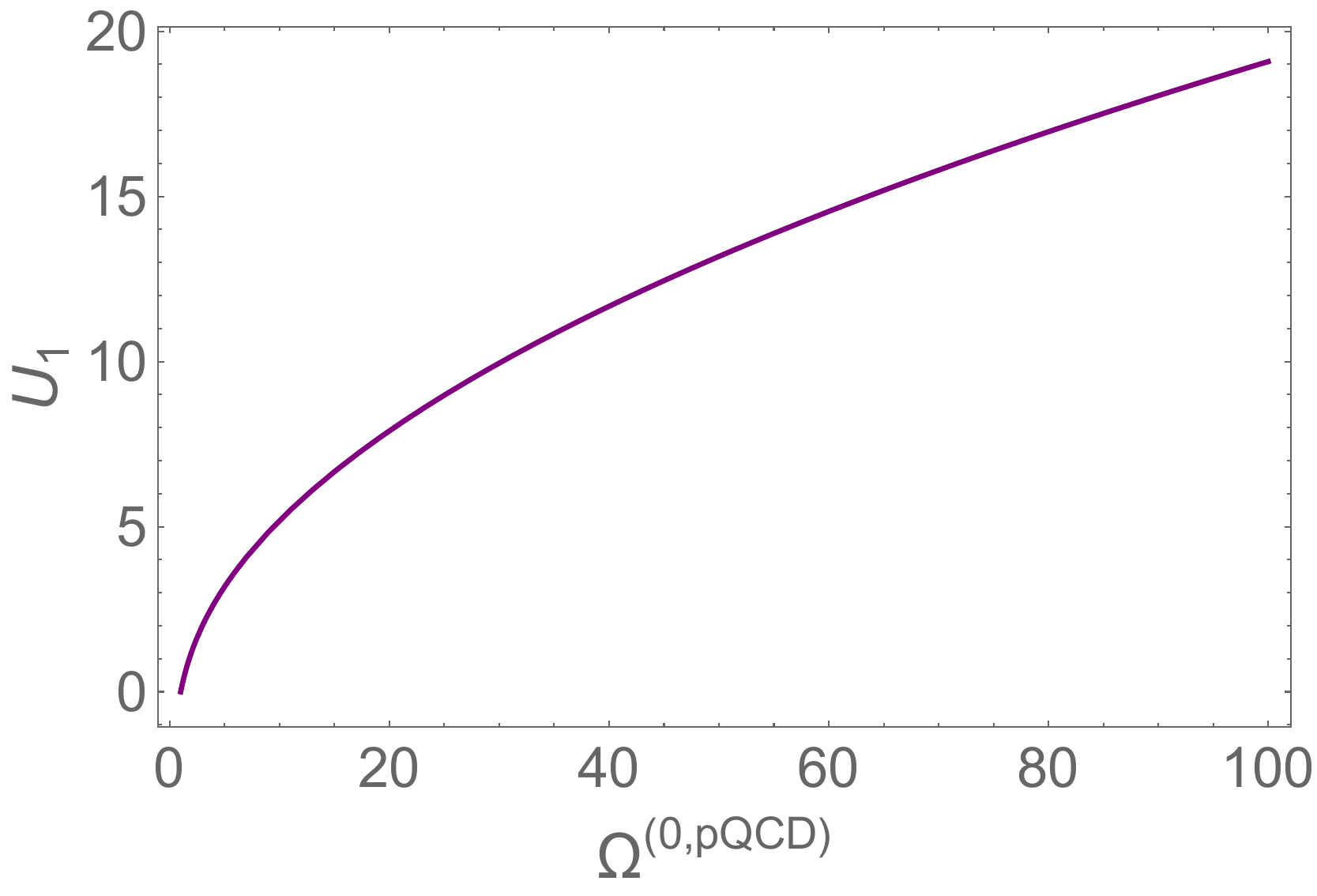}&\includegraphics[width=7.5cm]{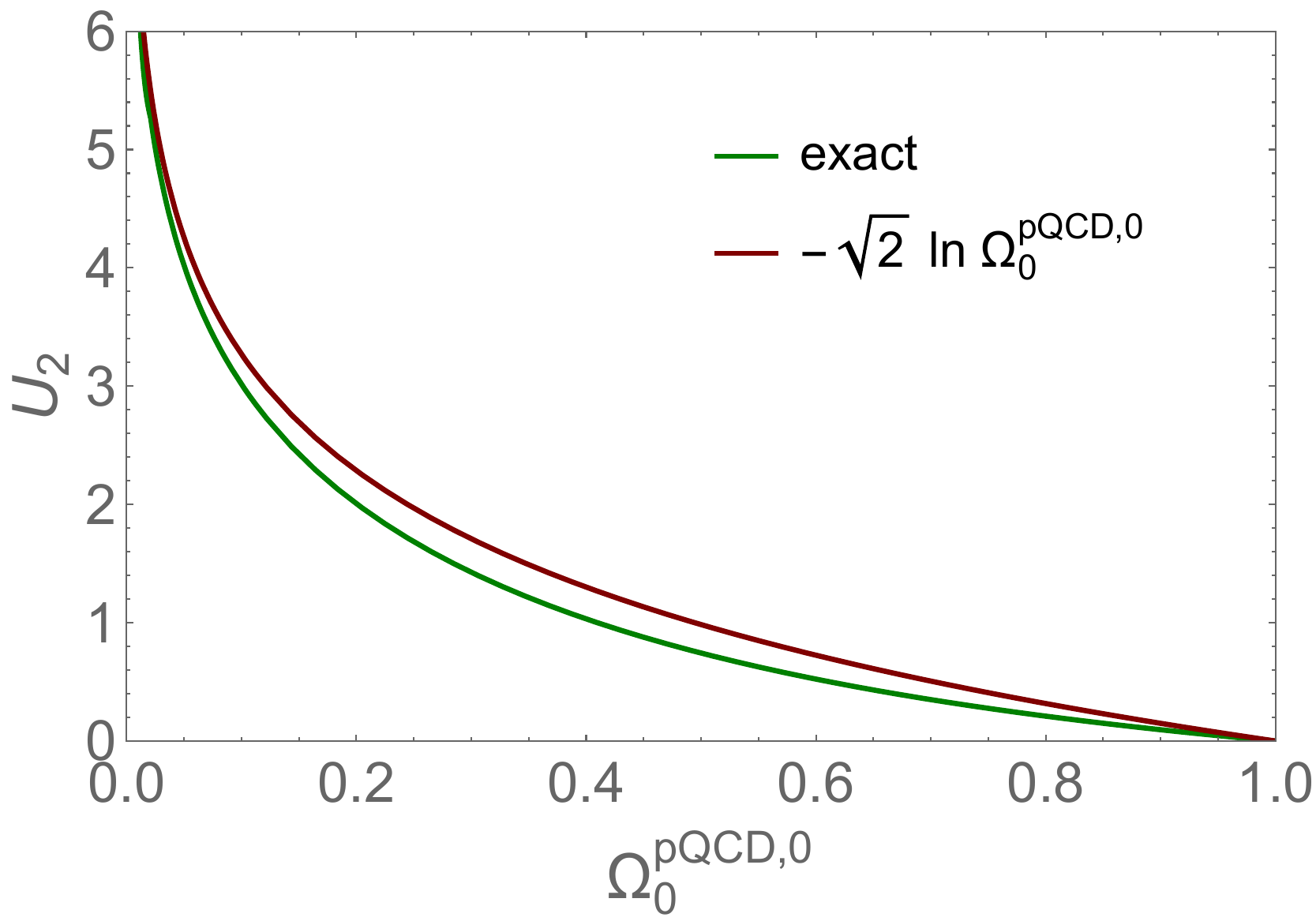} \\
		\fig{pu1}-a & \fig{pu1}-b\\
		\end{tabular}
			\end{center}
 	\caption{\fig{pu1}-a: Function  $\mathscr{U}_1\Lb\Omega^{0,pQCD}\Rb$ versus
	 $\Omega^{(pQCD,0)}$. \fig{pu1}-b: Function  $\mathscr{U}_2\Lb\Omega^{0,pQCD}_0\Rb$ versus
	 $\Omega^{0,pQCD}_0$}  
 	\label{pu1}
 \end{figure}
 
   In this equation we consider $\Omega^{0,pQCD}_0  \ll 1$   and it is  given by \eq{DIFPIC1}. Resulting $C_1$ is equal to\footnote{It is worthwhile mentioning that  in leading twist approximation for the BFKL kernel $C_1 = 0$ since $\bar{\gamma}=\h $  and $\kappa=4$\cite{LETU}.}
    \beq \label{DIFP5}   
   C_1\,\,=\,\,\Lb \bar{\gamma}^2 - \frac{1}{\kappa}\Rb\Lb \Omega^{0,pQCD}_0 \Rb^2
   \eeq              
       Finally, we have the implicit solution in the form (see \eq{GSS8} and \eq{GSS81}):
        \beq \label{DIFP6} 
  \intl^{\Omega^{0,pQCD}}_{\Omega^{0,pQCD}_0} \frac{ d \,\Omega'}{\sqrt{\Omega' +e^{-\Omega'} -1 + \h\Lb \bar{\gamma}^2\,\kappa\,-\,1\Rb\Lb\Omega^{0,pQCD}_0\Rb^2}}   =\sqrt{\frac{2}{\kappa}}\Lb z\,\,-\,\,z_0\Rb \,     \eeq       
  Using smallness of  $\Omega^{0,pQCD}_0$ we can rewrite \eq{DIFP6} as follows:
   \beq \label{DIFP6} 
 \underbrace{ \intl^{\Omega^{0,pQCD}}_{1} \frac{ d \,\Omega'}{\sqrt{\Omega' +e^{-\Omega'} -1 }}}_{\mathscr{U}_1\Lb\Omega^{0,pQCD}\Rb} \,\,+\,\,  
\underbrace{ \intl^{1}_{\Omega^{0,pQCD}_0} \frac{ d \,\Omega'}{\sqrt{\Omega' +e^{-\Omega'} -1 + \h\Lb \bar{\gamma}^2\,\kappa\,-\,1\Rb\Lb\Omega^{0, pQCD}_0\Rb^2}} }_{\mathscr{U}_2\Lb\Omega^{0,pQCD}_0\Rb} =\sqrt{\frac{2}{\kappa}}\Lb z\,\,-\,\,z_0\Rb \,   
 \eeq  
 where we neglect the $\Omega^{0,pQCD}_0$ contribution in  $\mathscr{U}_1$.
     
    The expression for  function $\mathscr{U}_1\Lb\Omega^{0,pQCD}\Rb$ it is more convenient to rewrite as follows:
   \beq \label{DIFP66}  
     \mathscr{U}_1\Lb\Omega^{0,pQCD}\Rb  \,\,=\,\,\intl^{\Omega^{0,pQCD} -1}_0    \frac{ d \,\Omega'}{\sqrt{\Omega' +e^{-\Omega' -1}}}     \eeq           
 \begin{figure}
 	\begin{center}
 	\leavevmode
 		\includegraphics[width=10cm]{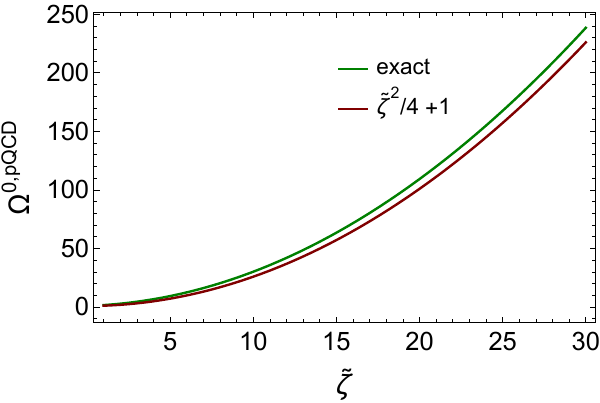}
		
			\end{center}
 	\caption{\fig{pu1}-a: $\Omega^{0,pQCD}$  versus $\tilde{\zeta} = \sqrt{\frac{2}{\kappa}}\Lb z - z_0\Rb - \mathscr{U}_2\Lb \Omega^{0,pQCD}_0\Rb$.} 	\label{pu2}
 \end{figure}          
             
      In \fig{pu1} we plotted  functions $\mathscr{U}_1$ and $\mathscr{U}_2$, while \fig{pu2} shows the solution for   $\Omega^{(pQCD,0)}$ for large and small values of $\zeta =\sqrt{\frac{2}{\kappa}}\Lb z - z_0\Rb - \mathscr{U}_2\Lb \Omega^{0,pQCD}_0\Rb$.

     Certainly,\eq{DIFP6} gives the solution only for $\Omega^{0,pQCD} \,>\,1$. For small  $\Omega^{0,pQCD} < 1$       we have       
  \beq \label{DIFP61} 
   \Omega^{0,pQCD}\,\,=\,\,\Omega^{0,pQCD}_0\Lb \cosh\Lb \frac{z - z_0}{\sqrt{\kappa}}\Rb \,+\,\bar{\gamma} \sqrt{\kappa} \sinh\Lb \frac{z - z_0}{\sqrt{\kappa}}\Rb \Rb
   \eeq   
                              
~
\subsection{First and second iterations  }
For the first iteration we can follow \eq{2I60} -\eq{2I7}, but introducing a new notation:
\beq \label{PQCDFI1}
z_D\,\,=\,\,\kappa (\tilde{Y} - \tilde{Y}_0)\,\,+\,\,\xi
\eeq
with $\xi\,\,=\,\,\ln\Lb Q^2_s\Lb \dy\Rb\,r^2\Rb$. $\Delta^D_0$  is given by \eq{DIFP6} (see \fig{pu2}). For estimates of the second iteration we take the simplified approach for $\Delta^D_0$, viz.:
\bea \label{PQCDFI2} 
 \Omega^{0,pQCD}\Lb \tilde{\zeta},\Omega^{0,pQCD}_0\Rb > 1 ~ & \Delta^{pQCD}_0\Lb z_D,\xi, z_0\Rb\,\,=&\,\,\exp\Lb - \Omega^{0,pQCD}\Lb \tilde{\zeta},\eq{DIFP6}\Rb\Rb \nn\\
 & \mbox{with}~~~ \tilde{\zeta} = &\sqrt{\frac{2}{\kappa}}\Lb z - z_0\Rb - \mathscr{U}_2\Lb \Omega^{(pQCD,0)}_0\Rb \nn\\
  \Omega^{0,pQCD}\Lb  \tilde{\zeta},\Omega^{0,pQCD}_0\Rb < 1 ~ & \Delta^{pQCD}_0\Lb z_D,\xi, z_0\Rb\,\,=& \Omega^{0,pQCD}_0\Lb \cosh\Lb \frac{z - z_0}{\sqrt{\kappa}}\Rb\,\,+\,\,\bar{\gamma} \sqrt{\kappa} \sinh\Lb \frac{z - z_0}{\sqrt{\kappa}}\Rb  \Rb
 \eea
  Considering $\Omega^{0,pQCD}_0$   being small we obtain for $\mathscr{U}_2\Lb \Omega^{0,pQCD}_0\Rb$:
   \beq \label{PQCDFI3}   
\mathscr{U}_2\Lb \Omega^{(pQCD,0)}_0\Rb = - \sqrt{2}\Lb \ln \Omega^{(pQCD,0)}_0 \,+\,\ln\Lb \bar{\gamma}^2\kappa - 1\Rb \, -\, \ln\frac{ 1 + \frac{1}{ \bar{\gamma}\,\sqrt{\kappa}}}
{1 - \frac{1}{ \bar{\gamma}\,\sqrt{\kappa}} }\Rb  \eeq
where   $ \Omega^{0,pQCD}_0\,\,=2\,N_0 \Lb r^2\, Q^2_s\Lb \dy\Rb\Rb^{ \bar{\gamma}}$  from \eq{DIFPIC1} .

\begin{figure}
 	\begin{center}
 	\leavevmode
	\begin{tabular}{c c}
 		\includegraphics[width=7cm]{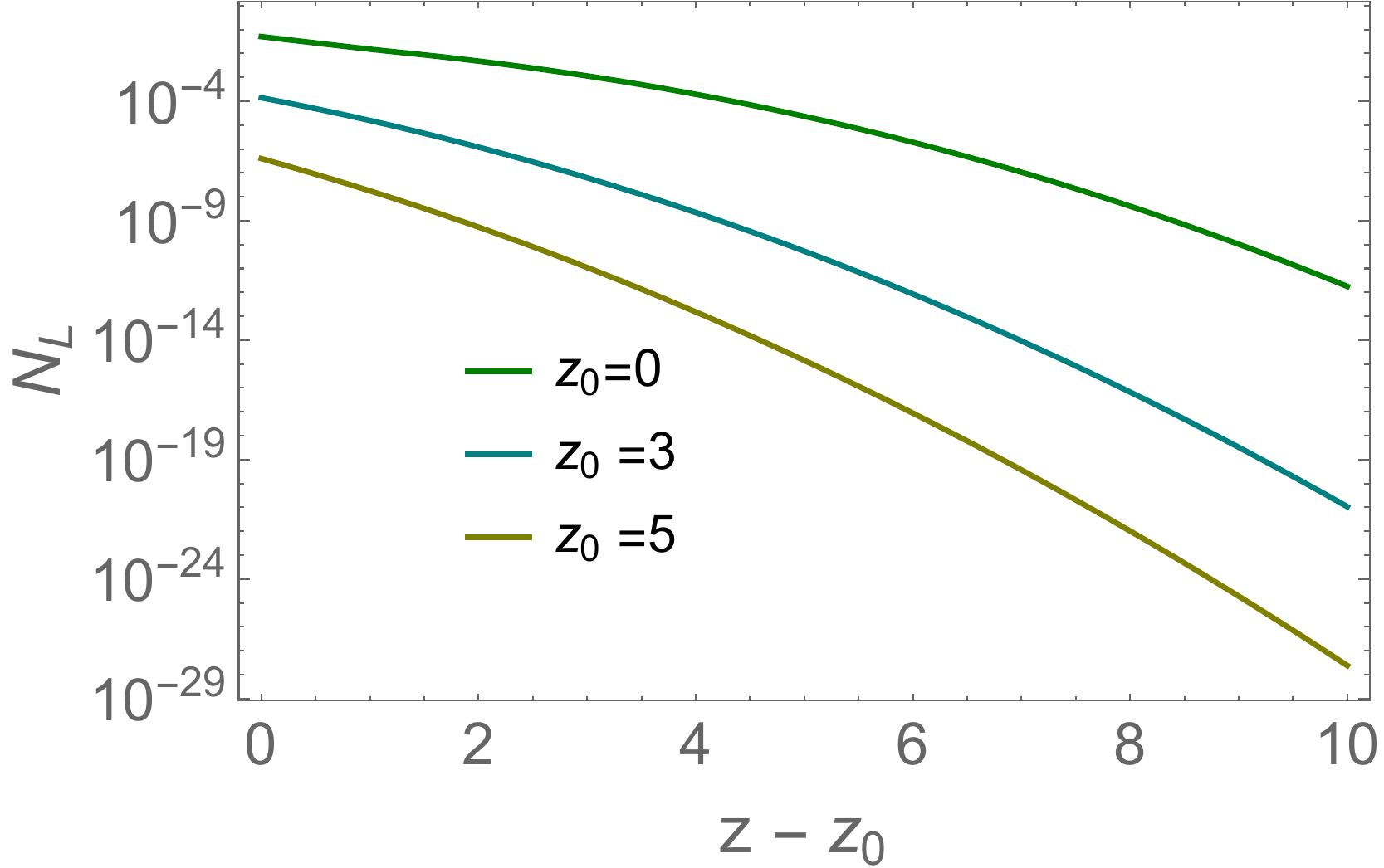}&\includegraphics[width=7cm]{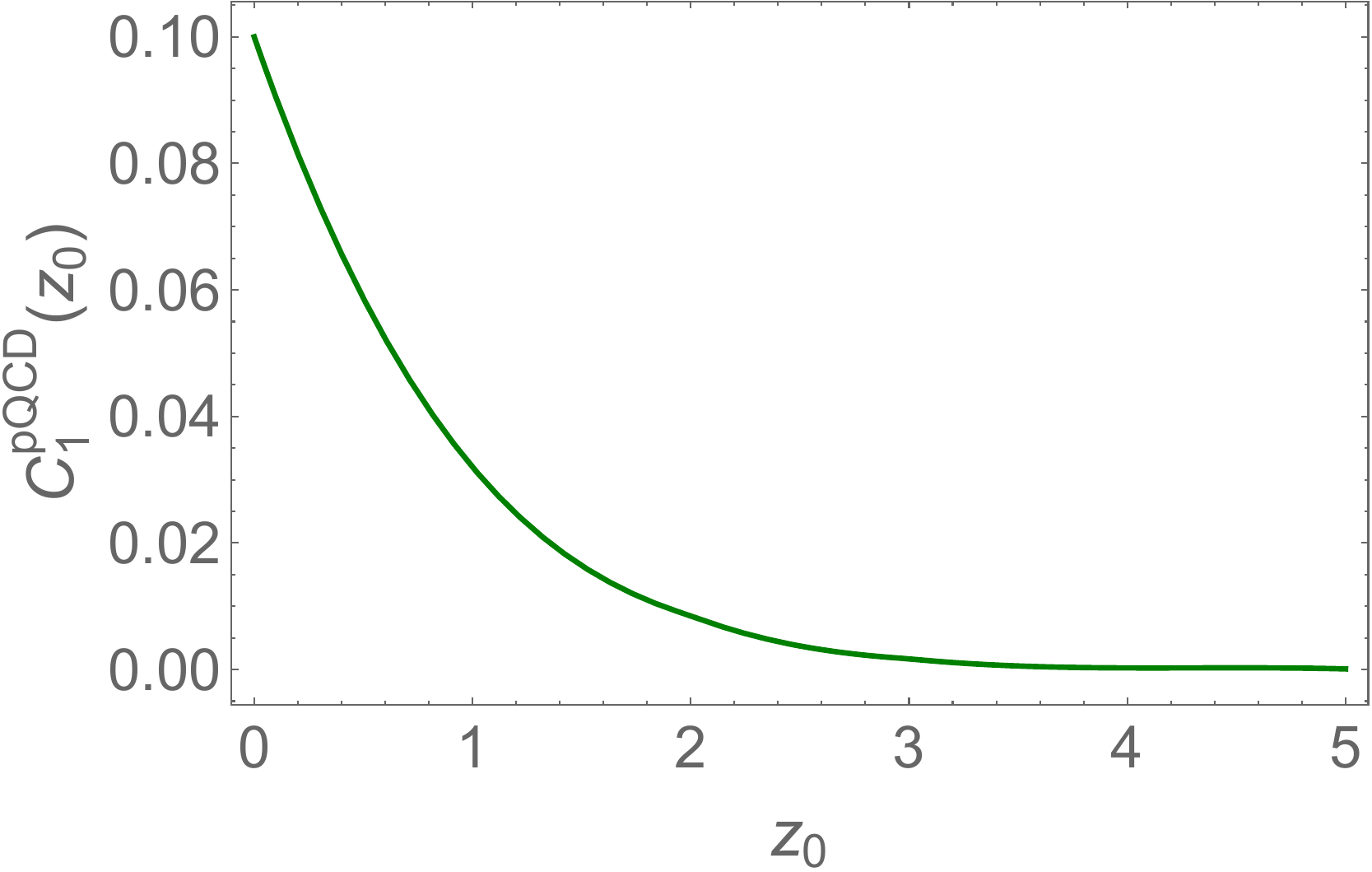}\\
		\fig{de1num}-a &\fig{de1num}-b\\
\end{tabular}
				\end{center}
 	\caption{\fig{de1num}-a: $\mathscr{N_L}$ versus $z$ at fixed $z_0$.
	\fig{de1num}-b: $C^{pQCD}_1$ versus $z_0$.
	 $N_0 = 0.25$ } 
\label{de1num}
 \end{figure}

       The general solution for the second iteration has the form (see \eq{2II1}):
     \beq \label{PQCDFI4}     
\Delta^D_1 \Lb z_D,\xi, z_0\Rb\,\,=\,\, \Delta^{pQCD}_0\Lb z_D,\xi, z_0\Rb\intl^{z_D}_{z_0} d z'  \frac{1}{\Delta^{pQCD}_0 \Lb z',\xi, z_0\Rb }\mathscr{N_L}
\eeq
   The contribution of  $\mathscr{N_L}$ we have discussed in section III-C  and it can be written as a sum of two contributions:
  \beq \label{PQCDFI5} 
 \mathscr{N_L} \,\,=\,\,-\underbrace{\Delta^{pQCD}_0\Lb z_D, z_0\Rb\intl_{z_0}^z  d z'  \Delta^{pQCD}_0\Lb z', z_0\Rb}_{\rm small \,\,size\,\,dipoles} 
+  \intl _{-z_0< \xi < \xi^A_0}\!\!\frac{d^2\,x_{02}}{2 \pi} \frac{ x^2_{01}}{x^2_{02}\,x^2_{12}} 
\Delta^D_0 \Lb z(\xi_{02}), z_0(\xi_{02})\Rb\Delta^D_0 \Lb z(\xi_{12}),  z_0(\xi_{12})\Rb
  \eeq 
  We divide the region of integration in two parts: (i) $z\Lb \xi_{02}\Rb < z$ and (ii) $z\Lb \xi_{02}\Rb > z$  . The first region cancels the first term in \eq{PQCDFI5}, while the second gives the main  contribution. In the second kinematic region we introduce $ \vec{x} =(x_1/r,x_2/r)$ where $\vec{x}_{02} = (x_1,x_2)$.  Using this notation  we replace $\frac{d^2\,x_{02}}{2 \pi} \frac{ x^2_{01}}{x^2_{02}\,x^2_{12}} $ by $\frac{dx_1\,d x_2}{(x^2_1 + x^2_2)\,((1-x_1)^2 + x^2_2)}$. In the  region $z\Lb \xi_{02}\Rb > z$ both
  $(x^2_1 + x^2_2) \geq 1$ and $ ((1-x_1)^2 + x^2_2)>1$.   
 In \fig{de1num}-a we show the result of integrations.

Since $ \Omega^{0,pQCD}_0$ is small we can hope that $\Delta^{pQCD}_1 \Lb z_D,\xi, z_0\Rb$ will be small as well (see \fig{d0d1}).
   Repeating the estimates in \eq{RIID17} - \eq{RIID18} we obtain
  
     \bea \label{PQCDFI7}     
\kappa \,\, \Delta^{pQCD}_1 \Lb z_D,\xi, z_0\Rb\,\,&=&\,\, \Delta^{pQCD}_0\Lb z_D,\xi, z_0\Rb\Bigg( \intl^{\infty}_{z_0} d z'  \frac{1}{\Delta^{pQCD}_0 \Lb z',\xi, z_0\Rb }\mathscr{N_L}
\,\, -\,\,\underbrace{\intl^{\infty}_{z_D} d z'  \frac{1}{\Delta^{pQCD}_0 \Lb z',\xi, z_0\Rb }\mathscr{N_L}}_{ \propto~~\frac{\kappa}{z} \Delta^{pQCD}_0\Lb z_D,\xi, z_0\Rb}\Bigg)\nn\\
&=&  \Delta^{pQCD}_0\Lb z_D,\xi, z_0\Rb\Bigg( {\rm C}^{\rm pQCD}_1 
\,\, -\,\,\intl^{\infty}_{z_D} d z'  \frac{1}{\Delta^{pQCD}_0 \Lb z',\xi, z_0\Rb }\mathscr{N_L}\Bigg)
\eea
     The second term in \eq{PQCDFI7}  decreases with $z_D$ and can be considered as a small correction. However the constant ${\rm C}^{\rm pQCD}_1$  has to be evaluated which are shown in  \fig{de1num}-b.
  One can see that  that the value of   ${\rm C}^{\rm pQCD}_1$  is rather small. Therefore, we have shown that our procedure of solving the nonlinear equation works.

~
\section{Conclusions }
In this paper we developed the homotopy approach for solving the non-linear evolution equation for the processes of the diffraction production in DIS\cite{KOLE}.  The goal of this approach is to choose  the first iteration by the simplified equation which can be solved analytically and includes the main features of the general solution. In our previous paper \cite{CLMNEW} for such equation the linearised version of the non-linear Balitsky-Kovchegov equation\cite{BK}  was used as the first iteration of the homotopy approach\cite{HE1,HE2}.  In this paper we introduced part of the nonlinear corrections in the first iteration (see \eq{SDBK12} for details). 

Based  on this approach we 
 found that   the first iteration of the homotopy  approach  gives the main contribution in the both kinematic regions which we consider  for the diffractive production: (i) $ r^2\,Q^2_s(Y) \,\geq\,1$ and $r^2Q^2_s( Y_0) \,\geq\,1$, where $Y_0 $ is the value of the rapidity gap; and (ii)  $ r^2\,Q^2_s(Y) \,\geq\,1$ and $r^2Q^2_s( Y_0) \,\leq\,1$. We also demonstrated that the second iteration of this approach leads to  small corrections. Therefore, we can conclude that we found the regular procedure of solving the non-linear equation in which we can take into account the small correction using the regular  perturbation  approximation.

 In the paper we focused mostly on
 the kinematic region $ r^2\,Q^2_s(Y) \,\geq\,1$ and $r^2Q^2_s( Y_0) \,\geq\,1$. We found that for $\xi \,<\,\xi^A_0$ (see region I in  \fig{sat}) our solution shows the geometric scaling behaviour,  while for $\xi \,>\,\xi^A_0$ (see region II in  \fig{sat})  this behaviour is strongly violated.   We found that the analytical solution of the non-linear equation that has been used as the first interaction, reproduces the intitial and boundary conditions in the both kinematic regions. The second iteration with zero initial and boundary conditions turns out to be small and could be taken into account together with higher iterations using the regular perturbative procedure.

For the kinematic region:  $ r^2\,Q^2_s(Y) \,\geq\,1$ and $r^2Q^2_s( Y_0) \,\leq\,1$ we found that the second iteration is small only f if $z_0 = \ln\Lb r^2 Q^2_s\Lb Y_0\Rb\Rb \,\geq \,0$. It means that for   the dipoles with very small sizes we could expect a considerable corrections in the perturbative approximation. We are going to investigate this region in our further publications.

~
\section{Acknowledgements}
We thank our colleagues at Tel Aviv University and UTFSM for encouraging discussions.  This research was supported by Fondecyt (Chile) grants No. 1231829 and  1231062.
J.G.  express his gratitude to the PhD scholarship USM-DP No. 029/2024 for the financial support.


\begin{thebibliography}{99} \frenchspacing
 
 \bibitem{CLMNEW}
C.~Contreras, E.~Levin and R.~Meneses,
Phys. Rev. D \textbf{107} (2023) no.9, 094030
doi:10.1103/PhysRevD.107.094030
[arXiv:2302.10497 [hep-ph]].

 \bibitem{HE1}
	J.H. He, 
	Comput. Methods Appl. Mech. Engrg. {\bf 173} (1999) 257.  
	
	\bibitem{HE2}
	J.H. He, 
	 Int. J. Nonlinear Mech. {\bf 35} (2000) 37.

\bibitem{BK}
I.~Balitsky,
{Phys.\ Rev.} {\bf D60}, 014020 (1999);
Y.~V.~Kovchegov,
{Phys.\ Rev.}  {\bf D60}, 034008  (1999).


  \bibitem{KOLE}
  Y.~V.~Kovchegov and E.~Levin,
  Nucl.\ Phys.\ B {\bf 577} (2000) 221.
  
    \bibitem{HWS} 
  M.~Hentschinski, H.~Weigert and A.~Schafer,
  Phys.\ Rev.\ D {\bf 73} (2006) 051501,
  [hep-ph/0509272].
   
   \bibitem{HIMST}
   Y.~Hatta, E.~Iancu, C.~Marquet, G.~Soyez and D.~N.~Triantafyllopoulos,
  Nucl.\ Phys.\ A {\bf 773} (2006) 95,
  [hep-ph/0601150].


\bibitem{KLW}
  A.~Kovner, M.~Lublinsky and H.~Weigert,
  Phys.\ Rev.\ D {\bf 74} (2006) 114023,
  [hep-ph/0608258].
     \bibitem{KLP}
A.~Kormilitzin, E.~Levin and A.~Prygarin,
Nucl. Phys. A \textbf{813} (2008), 1-13
doi:10.1016/j.nuclphysa.2008.09.006
[arXiv:0807.3413 [hep-ph]].
  
\bibitem{LEWU}
E.~Levin and M.~Wusthoff,
Phys. Rev. D \textbf{50} (1994), 4306-4327.
 \bibitem{GBKW}
 K.~J.~Golec-Biernat and J.~Kwiecinski,
  Phys.\ Lett.\ B {\bf 353} (1995) 329,
  [hep-ph/9504230].
  \bibitem{GOLEDD}
  E.~Gotsman, E.~Levin and U.~Maor,
  Nucl.\ Phys.\ B {\bf 493} (1997) 354,
  [hep-ph/9606280].
   
    \bibitem{SATMOD0}
     K.~J.~Golec-Biernat and M.~Wusthoff,
  Phys.\ Rev.\ D {\bf 60} (1999) 114023
  [hep-ph/9903358];\,\, 
  Phys.\ Rev.\ D {\bf 59} (1998) 014017;\,\,
  [hep-ph/9807513].
  
  \bibitem{KOML}
  Y.~V.~Kovchegov and L.~D.~McLerran,
  Phys.\ Rev.\ D {\bf 60} (1999) 054025
   Erratum: [Phys.\ Rev.\ D {\bf 62} (2000) 019901],
  [hep-ph/9903246].
       
  \bibitem{MUSCH}
  S.~Munier and A.~Shoshi,
  Phys.\ Rev.\ D {\bf 69} (2004) 074022,
  [hep-ph/0312022].
  
  \bibitem{MASC}
   C.~Marquet and L.~Schoeffel,
  Phys.\ Lett.\ B {\bf 639} (2006) 471,
  [hep-ph/0606079].
  \bibitem{MAR}
   C.~Marquet,
  Phys.\ Rev.\ D {\bf 76} (2007) 094017,
  [arXiv:0706.2682 [hep-ph]].
  
  
  \bibitem{KLMV}
  H.~Kowalski, T.~Lappi, C.~Marquet and R.~Venugopalan,
  Phys.\ Rev.\ C {\bf 78} (2008) 045201,
  [arXiv:0805.4071 [hep-ph]].
    
  
  \bibitem{LELUDD}
   E.~Levin and M.~Lublinsky,
  Nucl.\ Phys.\ A {\bf 712} (2002) 95,
  [hep-ph/0207374].
  \bibitem{LELUDD1}
   E.~Levin and M.~Lublinsky,
  Eur.\ Phys.\ J.\ C {\bf 22} (2002) 64,  [hep-ph/0108239].
 
 
   \bibitem{KOLEB}
Yuri V. Kovchegov and Eugene Levin, {\it `` Quantum Chromodynamics at High Energies"}, Cambridge Monographs on Particle Physics, Nuclear Physics and Cosmology, Cambridge University Press, 2012 .  

\bibitem{CLMP}
C.~Contreras, E.~Levin, R.~Meneses and I.~Potashnikova,
Eur. Phys. J. C \textbf{78} (2018) no.6, 475
doi:10.1140/epjc/s10052-018-5957-z
[arXiv:1802.06344 [hep-ph]].

\bibitem{CLMP1}
C.~Contreras, E.~Levin, R.~Meneses and I.~Potashnikova,
Eur. Phys. J. C \textbf{78} (2018) no.9, 699
doi:10.1140/epjc/s10052-018-6179-0
[arXiv:1806.10468 [hep-ph]].
\bibitem{MM1}
A.~H.~Mueller and S.~Munier,
Phys. Rev. D \textbf{98} (2018) no.3, 034021
doi:10.1103/PhysRevD.98.034021
[arXiv:1805.02847 [hep-ph]].
\bibitem{MM2}
A.~H.~Mueller and S.~Munier,
Phys. Rev. Lett. \textbf{121} (2018) no.8, 082001
doi:10.1103/PhysRevLett.121.082001
[arXiv:1805.09417 [hep-ph]].
\bibitem{MM3}
A.~D.~Le, A.~H.~Mueller and S.~Munier,
Phys. Rev. D \textbf{104} (2021), 034026
doi:10.1103/PhysRevD.104.034026
[arXiv:2103.10088 [hep-ph]].


 
 \bibitem{AGK}
V.~A.~Abramovsky, V.~N.~Gribov and O.~V.~Kancheli,
Yad. Fiz. \textbf{18} (1973), 595, (Sov.J. Nucl.Phys. 18 (1974),308);

 
 

   \bibitem{LIP}
 L.~N.~Lipatov,
  Sov.\ Phys.\ JETP {\bf 63}, 904 (1986)
  [Zh.\ Eksp.\ Teor.\ Fiz.\  {\bf 90}, 1536 (1986)].
 \bibitem{LIREV}
                L.~N.~Lipatov,
  Phys.\ Rept.\  {\bf 286} (1997) 131.
               

\bibitem{MUT}
A.~H.~Mueller and D.~N.~Triantafyllopoulos,
{\it Nucl.\ Phys.} \, {\bf B640} (2002) 331
[arXiv:hep-ph/0205167];\,\,D.~N.~Triantafyllopoulos,
{\it Nucl.\ Phys.}\,  {\bf B648} (2003) 293
[arXiv:hep-ph/0209121].

\bibitem{BFKL}
   V.~S. Fadin, E.~A. Kuraev and L.~N. Lipatov,
\newblock Phys. Lett. {\bf B60}, 50 (1975);\,\,\,
E.~A. Kuraev, L.~N. Lipatov and V.~S. Fadin,
\newblock Sov. Phys. JETP {\bf 45}, 199 (1977),
\newblock [Zh. Eksp. Teor. Fiz.72,377(1977)];\,\,\,
I.~I. Balitsky and L.~N. Lipatov,
\newblock Sov. J. Nucl. Phys. {\bf 28}, 822 (1978),
\newblock [Yad. Fiz.28,1597(1978)].
          
\bibitem{LETU}
E.~Levin and K.~Tuchin,
  Nucl.\ Phys.\ B {\bf 573}, 833 (2000)
  [hep-ph/9908317];\,\,\,
  Nucl.\ Phys.\ A {\bf 691}, 779 (2001)
  [hep-ph/0012167]; 
   {\bf 693}, 787 (2001)
  [hep-ph/0101275].
                

  
\bibitem{BALE}
J.~Bartels and E.~Levin,
Nucl. Phys. B \textbf{387} (1992), 617-637
doi:10.1016/0550-3213(92)90209-T
\bibitem{GS}
A.~M.~Stasto, K.~J.~Golec-Biernat and J.~Kwiecinski,
Phys. Rev. Lett. \textbf{86} (2001), 596-599
doi:10.1103/PhysRevLett.86.596
[arXiv:hep-ph/0007192 [hep-ph]].
L.~McLerran and M.~Praszalowicz,
Acta Phys. Polon. B \textbf{42} (2011), 99-103
doi:10.5506/APhysPolB.42.99
[arXiv:1011.3403 [hep-ph]];
Phys. Lett. B \textbf{741} (2015), 246-251
doi:10.1016/j.physletb.2014.12.046
[arXiv:1407.6687 [hep-ph]].

 \bibitem{GOST}
K.~J.~Golec-Biernat and A.~M.~Stasto,
Nucl. Phys. B \textbf{668} (2003), 345-363
[arXiv:hep-ph/0306279 [hep-ph]].
\bibitem{BEST}
J.~Berger and A.~Stasto,
Phys. Rev. D \textbf{83} (2011), 034015
[arXiv:1010.0671 [hep-ph]].

\bibitem{CLMS}
C.~Contreras, E.~Levin, R.~Meneses and M.~Sanhueza,
Eur. Phys. J. C \textbf{80}, no.11, 1029 (2020)
doi:10.1140/epjc/s10052-020-08580-w
[arXiv:2007.06214 [hep-ph]].

\bibitem{GLR} 
L.~V.~Gribov, E.~M.~Levin and M.~G.~Ryskin,
  Phys.\ Rept.\  {\bf 100}, 1 (1983).
 
    \bibitem{MATH2}
  A.D. Polyanin and V.F. Zaitsev,{\it ```Handbook of nonlinear partial differential equations"},CRC press,2004.   
\bibitem{MATH1}
  A.D. Polyanin and V.F. Zaitsev,{\it ``Handbook of Exact Solutions for Ordinary Differential Equation''},CRC press,1995. 

\bibitem{POLY}
Andrey D. Polyanin, ``Handbook of Linear Differential Equations For Engineers and Scientists", Chapman\& Hall/CRC,
2002.  
\bibitem{OLVER2010}
 F.Olver, D. Lozier, R.Boisvert and C. Clark, eds.,{\it `` NIST handbook of Mathematical Functions"}. U.S. Departament of Commerce, National Institute of Standard and Technology, Cambridge University Press, Wasington, DC, Cambridge, 2010.
 \bibitem{AS}
  M. Abramowitz and I. Stegun, {\it ``Handbook of Mathematical Functions with Formulas, Graphs, and Mathematical Tables''}, United States Department of Commerce, National Bureau of Standards,1964. 
 
   \end{thebibliography}
\end{document}